\documentclass[lineno]{JFM-FLM_Au}

\newcommand {\mb}{\boldsymbol}

\lefttitle{F. Daniel and D. Lecoanet}
\righttitle{Journal of Fluid Mechanics}

\title{A self-consistent numerical model of internal wave-induced mean flow oscillations in polar geometry}

\author{Florentin Daniel\aff{1} \and Daniel Lecoanet\aff{1,2}}

\affiliation{\aff{1}Center for Interdisciplinary Exploration and Research in Astrophysics (CIERA), Northwestern University, Evanston, IL, USA
\aff{2}Department of Engineering Sciences and Applied Mathematics,
Northwestern University, Evanston, IL, USA}

\corresau{F. Daniel, florentin.daniel@northwestern.edu}

\begin{document}
\nolinenumbers %for arvix
\maketitle

\begin{abstract}

%Building on the Direct Numerical Simulations of \citet{Couston2018}, we investigate the dynamics of a stably-stratified envelope and a convectively unstable core in polar geometry. As the coupling between the two zones is achieved self-consistently, internal gravity waves (IGW) generated by convection lead to the formation of a large scale oscillating azimuthal flow in the upper layer. The emphasis is put on characterising its period and typical velocity, as well as its regularity. Despite a continuous broad spectrum of IGW, our work show qualitative good agreement with the monochromatic model of \citet{Plumb1978}. If the latter was originally developed in the context of the Earth’s quasi biennial oscillation, our study could prove relevant for its stellar counterpart in massive stars, which host a convective core and a radiative envelope.
The Earth’s Quasi-Biennial Oscillation (QBO) is a natural example of wave-mean flow interaction and corresponds to the alternating directions of winds in the equatorial stratosphere. It is due to internal gravity waves (IGW) generated in the underlying convective troposphere. In stars, a similar situation is predicted to occur, with the interaction of a stably-stratified radiative zone and a convective zone. In this context, we investigate the dynamics of this reversing mean flow by modelling a stably-stratified envelope and a convectively unstable core in polar geometry. Here, the coupling between the two zones is achieved self-consistently, and IGW generated through convection lead to the formation of a reversing azimuthal mean flow in the upper layer. We characterise the mean-flow oscillations by their periods, velocity amplitudes, and regularity. Despite a continuous broad spectrum of IGW, our work show good qualitative agreement with the monochromatic model of \citet{Plumb1978}. If the latter was originally developed in the context of the Earth’s QBO, our study could prove relevant for its stellar counterpart in massive stars, which host convective cores and radiative envelopes.

\end{abstract}

%\begin{keywords}
%bla
%\end{keywords}

%{\bf MSC Codes }  {\it(Optional)} Please enter your MSC Codes here

\section{Introduction}
\label{sec:intro}

Geophysical and Astrophysical Fluid Dynamics (GAFD) exhibits numerous examples where strong spatial and temporal separation of scales lead to rich nonlinear behaviours. Among them, the reversals of winds between eastern and western directions in the Earth's stratosphere is one of the most striking. Known as the Quasi-Biennial Oscillation (QBO), it was first quantitatively measured in the early 1960s \citep{Ebdon1960, Reed1961}. Due to the internal gravity waves (IGW) induced by convective motions in the underlying troposphere, a downward pattern of zonal winds is generated through the alternating deposit of angular momentum by prograde and retrograde waves. Subsequent measurements showed the period of the flow to be of 28 months in average, leading to its name \citep{Baldwin2001}. It is much longer than the typical period of the waves which is of a few days.

Earth is not the only illustration of such a phenomenon. Jupiter also displays similar reversals, the Quasi-Quadrennial Oscillation with a period of approximately 54 months \citep{Leovy1991}, and Saturn with a period of 14.8 years, where it is related to the planet seasonal forcing \citep{Orton2008}. The three planetary oscillations mentioned here are all equatorially confined.

In stars, there is a similar situation necessary for QBO, with a stably-stratified layer---the radiative zone (RZ)---which is adjacent to a convectively unstable one---the convective zone (CZ). This has led the community to posit the existence of the same type of behaviour, known as shear-layer oscillations \citep{Kumar1999, Charbonnel2005, Fuller2014}. However, such shear-layer oscillations have yet to be observed.

While wave-driven mean flow oscillations in GAFD arise from the interplay between convective motions and stably-stratified layers, early works managed to capture the oscillations only modelling the stably-stratified layer. \citet{Plumb1977} and \citet{Plumb1978} introduced a one-dimensional model accounting for the evolution of the mean flow excited by two monochromatic waves with opposite phase velocities (see § \ref{sec:theory} for more details). Their work, which was corroborated in a laboratory experiment, sparked considerable interest in the atmospheric sciences community, as the model proved to be able to reproduce many qualitative features of the QBO \citep[e.g.][]{Wedi2006,Renaud2019}, being besides a source of rich nonlinear behaviours. \citet{Yoden1988} indeed showed, through numerical integration of the Plumb and McEwan model, that the mean flow follows the behaviour of a Hopf bifurcation when the amplitude of the waves is changed. This was later confirmed by \citet{Semin2018} via a weakly non linear analysis of the system \citep[see also][]{Semin2024}. Transitions to chaos were also found \citep{Kim2001}.

While these one-dimensional models have had success in representing certain aspects of the QBO, many researchers have developed more complex models to address new observations of QBO disruptions, as well as make predictions of similar behaviours in other planets and stars. In the atmospheric sciences community, \citet{Saravanan1990} for instance took into account several waves (see also \citet{Leard2020} or \citet{Chartrand2024} more recently). Stochastic wave excitation has also been considered \citep{Ewetola2024}.

In the astrophysical community, the one-dimensional models have been applied to stellar parameters \citep{Talon2002, Talon2003, Charbonnel2005}, and have also been extended to include additional physical effects. \citet{Kim2003} for instance included the influence of magnetic fields and applied their calculations to the Sun. The sophistication of Direct Numerical Simulations (DNS) led to the development of global models, where contrary to Plumb and McEwan, the action of the CZ is resolved self-consistently and not through any form of parameterisation. \tck{For solar-like stars, where the RZ is below the CZ, \citet{Rogers2006} argued that hints of a mean-flow reversal could be seen in their simulations.} In the context of massive stars, \tck{where the location of the two zones is inverted,} \citet{Rogers2012} and \citet{Rogers2013} reported the development of a mean flow in the RZ. The authors attributed its development to the continuous spectrum of IGW generated by the CZ, but did not observe any reversals.

%In the astrophysical community, the lack of observational constraints means that no direct comparisons can be conducted with the reduced model. If the latter has been applied to stellar parameters \citep{Talon2002, Talon2003}, the many uncertainties about the relevant physical ingredients to be considered led to some modification of Plumb and McEwan's model. The inclusion of magnetic field by \citet{Kim2003} was directly applied to the case of the Sun. It thus followed earlier complexifications of Plumb and McEwan's model in the atmospheric sciences community. \citet{Saravanan1990} for instance took into account several waves (see also \citet{Leard2020} or \citet{Chartrand2024} more recently). It is the sophistication of Direct Numerical Simulations (DNS) that led to the development of global models, where contrary to Plumb and McEwan, the action of the CZ is resolved self-consistently and not through any form of parameterisation. In the context of massive stars, \citet{Rogers2012} and \citet{Rogers2013} reported the development of a mean flow in the RZ. The authors attributed its development to the continuous spectrum of IGW generated by the CZ, but did not observe any reversals.

The first study to indeed observe an oscillating mean flow while explicitly solving for the dynamics of the two zones was conducted by \citet{Couston2018} in a 2D Cartesian geometry. Adopting a model where the thermal expansion coefficient of the modelled fluid differs in the two zones (see \citet{Couston2017} and § \ref{sec:method}), they were able to generate periodic reversals of the mean flow. Their work highlighted the key role of the ratio of molecular viscosity to thermal diffusivity in order to favour oscillations.

While the initial motivation of \citet{Couston2018} came from the context of laboratory experiments where the nonlinear equation of state of fresh water allows for such mixed layer dynamics, it is the goal of the present paper to extend their model to polar geometry, as a model of the equatorial slice of a star. Being able to thoroughly understand the complex interplay between the two zones with DNS is of the utmost importance to motivate future closure models for stellar evolution codes. A direct stellar application could be for instance to compare the CZ statistics of our work to the parameterised one used by \citet{Showman2019} for Brown Dwarfs oscillations. A better understanding of the mechanism behind wave-driven mean flow oscillations can also be applied to other geophysical and astrophysical contexts (see \citealt{Bardet2022} for the case of Saturn).

The organisation of this paper is as follows. We first introduce our numerical model in § \ref{sec:method}. As our numerical model has many similarities to the Plumb and McEwan model, we recall in § \ref{sec:theory} the main characteristics of their work for completeness, adapted to our setup, and refer to the original study. We present results in § \ref{sec:results}. We detail the waves generated by the CZ (§ \ref{sec:res:waves}) and then present properties of the mean flow---its period and typical velocity---in § \ref{sec:res:meanflow}. Implications of our study as well as directions for future works are discussed in § \ref{sec:ccl}. %Further technical precisions on our numerical model are given in Appendix \ref{appA}, related to the location of the interface between the two zones. A table of the simulations is available in Appendix \ref{appB}.

\section{Methods}\label{sec:method}

\begin{figure}[t]
  \centerline{\includegraphics[width=14cm]{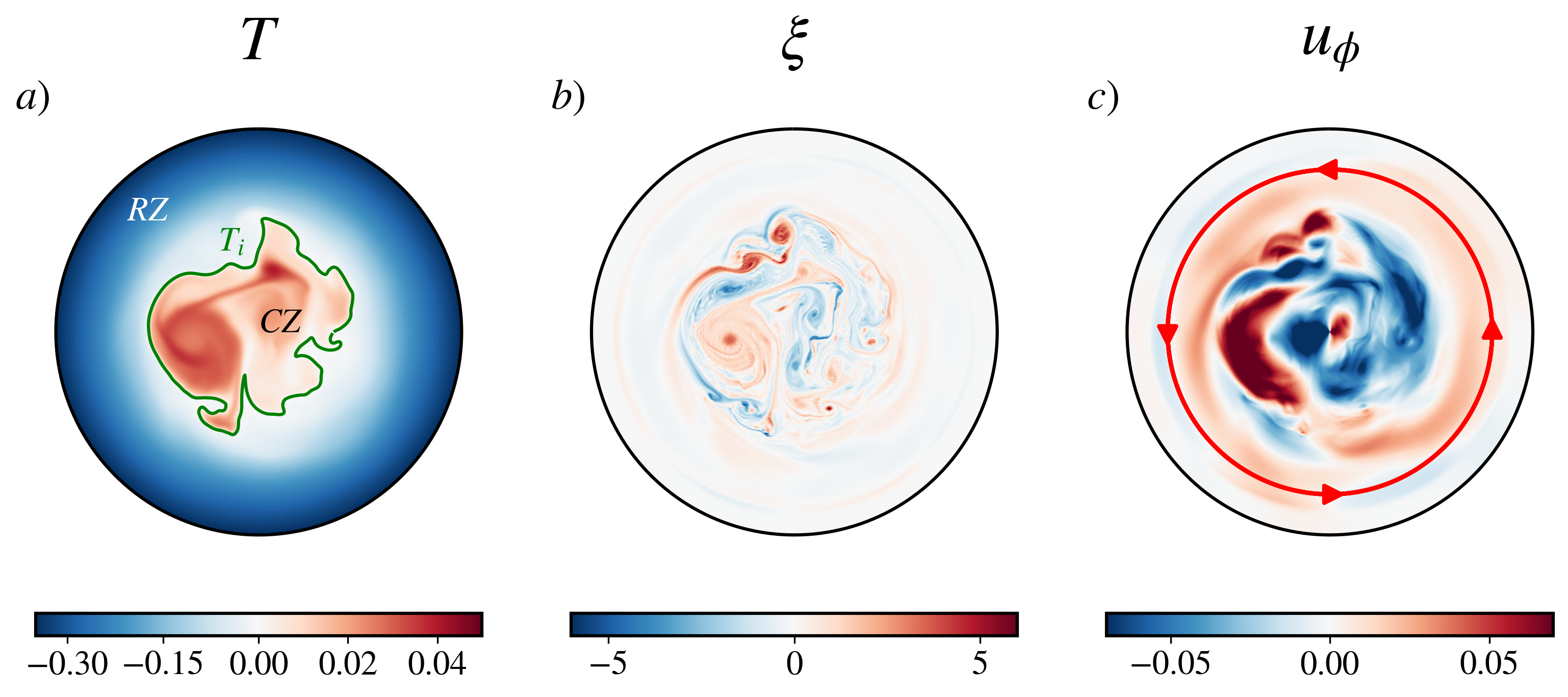}}% Images in 100% size
  \caption{Visualisations of convection, waves, and mean flows in the problem, in code units (see § \ref{sec:method}). \textit{a)} Temperature field in a simulation, displaying an active core compared to the envelope. The contour value $T_i=0$ denotes the boundary between the convective (CZ) and radiative (RZ) zones. The colorbar is stretched, going from $-0.35$ (blue) to $0$ (white) in the RZ, and from $0$ to $0.05$ (red) in the CZ. \textit{b)} Vorticity $\xi=\nabla \times \mb{u}$ of the flow, illustrating the typical turbulence generated in the core of the domain. \textit{c)} Azimuthal flow developing in the stably-stratified layer, here showing a counter clockwise mean component whose reversal is at the core of the present study.}
\label{fig:1}
\end{figure}

We study a 2D non-rotating fluid in a disk of radius $\widetilde{r_o}$. The fluid is governed by the Navier-Stokes equations:

\begin{align}
	\frac{\partial \widetilde{\mb{u}}}{\partial \widetilde{t}} + \mb{\widetilde{u}} \cdot \widetilde{\nabla} \mb{\widetilde{u}}  &= -  \frac{ \widetilde{\nabla} \widetilde{p}}{\widetilde{\rho_0}} +  \widetilde{\nu} \widetilde{\Delta} \widetilde{\mb{u}} - \frac{\widetilde{\delta \rho}}{\widetilde{\rho_0}}(\widetilde{T})\widetilde{g}(\widetilde{r}\,)\mb{e_r} - \widetilde{N} D(\widetilde{r}\,)\widetilde{\mb{u}}, \label{eq_U_dim}\\
	\frac{\partial \widetilde{T}}{\partial \widetilde{t}} +  \widetilde{\mb{u}} \cdot \widetilde{\nabla} \widetilde{T} &= \widetilde{\kappa} \widetilde{\Delta}\, \widetilde{T} + \widetilde{Q}(\widetilde{r}\,), \label{eq_THETA_dim}\\
	\widetilde{\nabla} \cdot \widetilde{\mb{u}} &= 0, \label{eq_DIVU_dim}
\end{align}

\noindent evolving $\widetilde{\mb{u}}$, $\widetilde{T}$ and $\widetilde{p}$, respectively the velocity, temperature and pressure of the fluid, reported here in their dimensional forms with tildes. $\widetilde{\nu}$ and $\widetilde{\kappa}$ are the kinematic viscosity and thermal diffusivity, assumed to be constant throughout the domain. The gravity profile is linear in the radius $\widetilde{g}(\widetilde{r}\,)=\widetilde{g_o}\,\widetilde{r}/\widetilde{r_o}$ with $\widetilde{g_o}$ its value at the surface of the domain. \tck{We now present all the assumptions and notations of (\ref{eq_U_dim}-\ref{eq_DIVU_dim}).}

We model the mixed dynamics of a stably-stratified envelope and a convectively unstable core, similar to the geometry of a massive star. Several methods exist in order to account for the coupled dynamics of both a convective zone (CZ) and a radiative zone (RZ). For example, one can impose a background density gradient changing sign at the interface between the two regions (see, e.g., \citet{Aubert2025} in the geodynamo context). Here, we follow the description of \citet{Couston2017}. Originally motivated by the behaviour of water whose density peaks at $4^\circ \,{\rm C}$ in laboratory conditions (see for instance \citet{Vallis2017}), we impose that the density variations be piecewise linear as a function of temperature,
\begin{equation}
\frac{\widetilde{\delta \rho}}{\widetilde{\rho_0}}(\widetilde{T}) =\left\{
\begin{aligned}
   S &\widetilde{\alpha_{CZ}} \widetilde{T}, \quad  \widetilde{T}\le \widetilde{T_i} , \\
   - &\widetilde{\alpha_{CZ}} \widetilde{T}, \quad   \widetilde{T}\ge \widetilde{T_i}.
\end{aligned}
\right.
\label{eq_density}
\end{equation}
In line with the typical Boussinesq approximation, we only take into account the density variations $\widetilde{\delta\rho}$ with respect to the background density of the fluid $\widetilde{\rho_0}$ in the buoyancy force. Equation (\ref{eq_density}) models the associated change of sign of the thermal expansion coefficient of the fluid, leading to two different behaviours: below the inversion temperature, $\widetilde{T_i}$, the flow is stably stratified, and above the inversion temperature the flow is convectively unstable. In the stellar context, this simulates a change from heat transfer by thermal convection in the core to heat transfer by radiative diffusion of photons in the envelope. $S$ is the stiffness parameter controlling the amplitude of stable stratification of the RZ via the Brunt-Väisälä frequency

\begin{equation}
    \widetilde{N}^2=-S\widetilde{\alpha_{CZ}}\widetilde{g}(\widetilde{r}\,) \frac{\partial \widetilde{T_{RZ}}}{\partial \widetilde{r}},
    \label{eq_N}
\end{equation}
\noindent where $\widetilde{\alpha_{CZ}}$ is the thermal expansion coefficient of the CZ. This is illustrated in Figure \ref{fig:1}a, where the core, whose temperature is above $\widetilde{T_i}$ is more active than the envelope. High values of $S$ tend to circularise the interface, as convective plumes are prevented from penetrating the strongly-stratified medium \citep{Couston2017}.

Convection is triggered through a volumetric heat source $ \widetilde{Q}(\widetilde{r}\,)=\widetilde{Q_0} e^{-(\widetilde{r}/\widetilde{r_b})^2}$. Its Gaussian shape peaking in the centre represents the action of nuclear reactions inside the star, similar to earlier studies in the same geometry \citep{Rogers2013}. In what follows, we set $ \widetilde{r_b}=0.1 \widetilde{r_o}$. While varying $\widetilde{r_b}$ would undeniably change the properties of the CZ quantitatively, we have made sure to choose it small compared to the interface radius $\widetilde{r_i}$ (see after and Appendix \ref{appA}), in order for internal heating and turbulence to be contained in the CZ only (see Figure \ref{fig:1}b). It is their action that generates the propagation of internal gravity waves (IGW) in the RZ. We include a damping layer in the outer portion of the domain to prevent wave reflection with profile $D(\widetilde{r}\,)=\left( 1 + \tanh ((\widetilde{r}-\widetilde{r_d})/\widetilde{\delta})\right)/2$. Following \citet{Couston2018}, we work with $\widetilde{r_d}=0.9\widetilde{r_o}$, $\widetilde{\delta}=0.02\widetilde{r_o}$. We discuss the properties of this damping layer in § \ref{sec:ccl}.

%Introducing $T_0=T_s(0)-T_s(r_i)$, $r_o$, $t_0=\left( r_o/\alpha_{CZ} g_o T_0\right)^{1/2}$ and $p_0=\rho_0 r_o^2/t_0^2$ respectively as temperature, length, time and pressure units, with  $T_s$ the static temperature profile, solution of $\Delta T = -Q(r)/\kappa$ (see Appendix \ref{appA}), every variable is written as $X = \widetilde{X}X_0$, with $X_0$ its unit. 
We non-dimensionalise the equations with $\widetilde{T_0}=\widetilde{T_s}(0)-\widetilde{T_s}(\widetilde{r_i^s}\,)$, $\widetilde{r_o}$, $\widetilde{t_0}=\left( \widetilde{r_o}/\widetilde{\alpha_{CZ}} \widetilde{g_o} \widetilde{T_0}\right)^{1/2}$ and $\widetilde{p_0}=\widetilde{\rho_0} \widetilde{r_o}^2/\widetilde{t_0}^2$ respectively as temperature, length, time, and pressure scales, and write each variable as $\widetilde{X}=\widetilde{X_0}X$ with $X$ dimensionless. Here, $\widetilde{T_s}$ is the conductive temperature profile, the solution to $\widetilde{\Delta}\, \widetilde{T} = -\widetilde{Q}(\widetilde{r}\,)/\widetilde{\kappa}$, and $\widetilde{r_i^s}$ the interface radius in this static case (see Appendix \ref{appA}). Equations (\ref{eq_U_dim}-\ref{eq_DIVU_dim}) become: 

%\begin{align}
	%\frac{\partial \mb{u}}{\partial t} + \mb{u} \cdot \nabla \mb{u}  &= - \nabla p +  \left(\frac{Pr}{Ra}\right)^{1/2} \Delta \mb{u} - f (T) r T \mb{e_r} - Nt_0 D(r)\mb{u}, \label{eq_U}\\
	%\frac{\partial T}{\partial t} +  \mb{u} \cdot \nabla T &= \frac{1}{\left(Ra Pr \right)^{1/2}} \left(\Delta T + \left(1-\frac{T_o}{T_0} \right) I_{r_b} e^{-(r/r_b)^2}\right), \label{eq_THETA}\\
%	\nabla \cdot \mb{u} &= 0. \label{eq_DIVU}
%\end{align}

\begin{align}
	\frac{\partial \mb{{u}}}{\partial {t}} + \mb{{u}} \cdot {\nabla} \mb{{u}}  &= - {\nabla} {p} +  \left(\frac{Pr}{Ra}\right)^{1/2} {\Delta} \mb{{u}} - f ({T}) {r} \,{T} \mb{e_r} - ND(r)\mb{{u}}, \label{eq_U}\\
	\frac{\partial {T}}{\partial {t}} +  \mb{{u}} \cdot {\nabla} {T} &= \frac{1}{\left(Ra Pr \right)^{1/2}} \left({\Delta}\, {T} + \left(1-T_o \right) I_{r_b} e^{-(r/r_b)^2}\right), \label{eq_THETA}\\
	{\nabla} \cdot \mb{{u}} &= 0. \label{eq_DIVU}
\end{align}

The temperature is expressed as the offset from $\widetilde{T_i}$, so that $T_i=0$ in our units. Equation (\ref{eq_density}) is rewritten as $f=(\widetilde{\delta \rho}/\widetilde{\rho_0})/\widetilde{\alpha_{CZ}}\widetilde{T}$. The integral $I_{r_b}$ depends only on the geometry of the internal heating profile, and is expressed in Appendix \ref{appA}. The dimensionless parameters controlling the behaviour of (\ref{eq_U}-\ref{eq_DIVU}) are
\begin{equation}
	Ra= \frac{\widetilde{\alpha_{CZ}}\widetilde{g_o} \widetilde{T_0} \widetilde{r_o}^3}{\widetilde{\nu}\,\widetilde{\kappa}},	\quad Pr = \frac{\widetilde{\nu}}{\widetilde{\kappa}}, \quad S, \quad T_o,
    \label{eq:paramDNS}
\end{equation}
%\stepcounter{equation} % Step first to reserve a new number
%\begin{equation}
%	Ra= \frac{\alpha_{CZ}g_o T_o r_o^3}{\nu\kappa},	\quad Pr = \frac{\nu}{\kappa}, \quad S, \quad \frac{T_o}{T_0},
 %   \tag{\theequation a--d}  \label{eq:paramDNS}
%\end{equation}
%\stepcounter{equation} % Step first to reserve a new number
\noindent where $Ra$ and $Pr$ are the Rayleigh and Prandtl numbers respectively. $S$ controls the value of the Brunt-Väisäla frequency $N=\sqrt{ST_o/\ln(r_i)}$, which with the polar geometry and profile of $g$ is constant. At the outer boundary, we impose stress-free and $T(1)=T_o$ Dirichlet temperature boundary conditions.

In this mixed-layer setup, the interface between the two zones $r_i$ is not an input parameter, contrary to studies where a background adiabatic gradient is imposed. Rather, $r_i$ is controlled by the temperature boundary condition, as we show in Appendix \ref{appA}, as the thermal flux in the CZ and RZ must match \tck{when taking a time average}. Here we take $T_o=-0.35$, enforcing $r_i\approx 0.5$. Similarly, we take $Pr=0.01$, motivated by earlier work suggesting that low $Pr$ systems tend to favour reversals (\citet{Couston2018}). These occur when IGW generated by the turbulent core deposit and extract angular momentum in the stably-stratified layer, leading to the development of a mean azimuthal flow in the RZ (Figure \ref{fig:1}c).

We integrate (\ref{eq_U}-\ref{eq_DIVU}) using the pseudo-spectral code Dedalus \citep{Burns2020}, setting two different grids in the radial direction in order to refine the discretisation near the interface between the CZ and RZ \citep{Vasil2016}. Concretely, we have a first disk domain with $N_{r1}$ radial points from $r=0$ to $r=0.6$, and a second annular domain with $N_{r2}$ radial points from $r=0.6$ to $r=1$. The azimuthal discretisation is achieved through Fourier series. As the typical oscillation of the stably-stratified layer is large compared to the convective turnover time, this study required long time integrations in order to obtain significant temporal statistics, of the order of several thermal diffusive times $\tau_\kappa=\sqrt{RaPr}$. The temporal evolution is conducted using a 2-step implicit/explicit Runge-Kutta time scheme, using a CFL condition with safety factor of $0.4$ \citep{Ascher1997}. Typical time steps varied between $10^{-4}$ and $10^{-3}$. Details of the simulations are given in Appendix \ref{appB}.

\section{Theoretical considerations}\label{sec:theory}

As mentioned in § \ref{sec:intro}, the first studies of wave-induced mean-flow oscillations simplified the problem by considering externally forced waves in a stably-stratified medium. Here we review this approach and apply it to our setup, as we use it to interpret our results (§ \ref{sec:results}). %We refer to 

%Considering a Cartesian model periodic in the horizontal direction and semi-infinite in the vertical ($z$), the horizontal component of (\ref{eq_U_dim}) once averaged $\langle .. \rangle $ in the zonal direction reads:

Consider a horizontally-periodic, vertically semi-infinite Cartesian domain, corresponding, respectively, to the zonal and radial directions of our polar geometry. At the base of this stably-stratified layer ($z=0$), two waves of equal amplitudes are generated with same frequencies $\widetilde{\omega}$ and horizontal wavenumbers $\widetilde{k}$ but opposite phase velocities $\pm \widetilde{c}$. Following \citet{Plumb1978}, we zonally average $\langle .. \rangle$ the zonal velocity equation (\ref{eq_U_dim}) in the bulk of the RZ, i.e. assuming $D=0$, and derive the dimensionless mean flow evolution equation
%\begin{equation}
 %   \frac{\partial \langle u \rangle }{\partial t} = -\frac{\partial \langle u' w'\rangle}{\partial z} + \Lambda_1 \frac{\partial^2 \langle u\rangle}{\partial z^2} - \Lambda_2 \langle u \rangle,
  %  \label{eq:Plumb}
%\end{equation}
\begin{equation}
    \frac{\partial \langle {u} \rangle }{\partial {t}} = -\frac{\partial \langle {u'} {w'}\rangle}{\partial {z}} + \Lambda_1 \frac{\partial^2 \langle {u}\rangle}{\partial {z}^2}.% - \Lambda_2 \langle {u} \rangle.
    \label{eq:Plumb}
\end{equation}
\tck{Considering low-frequency waves $\widetilde{\omega} \ll \widetilde{N}$ and neglecting cross-wave interactions, use of a Wentzel-Kramers-Brillouin (WKB) approximation can be made so we can write the Reynolds stress term as (e.g. \citet{Couston2018} supplementary material)}
%\begin{equation}
 %   \langle u' w' \rangle= \pm \exp{-\int_0^z \left( \frac{\alpha_1}{(\langle u\rangle\mp 1)^2} + \frac{\alpha_2}{(\langle u \rangle\mp 1)^4}\right)dz}.
  %  \label{eq:Reynolds}
%\end{equation}
\begin{equation}
    %\langle {u'} {w'} \rangle= \pm \exp{-\int_0^{{z}} \left( \frac{\alpha_1}{(\langle {u}\rangle\mp 1)^2} + \frac{\alpha_2}{(\langle {u} \rangle\mp 1)^4}\right)dz'}.
    \langle {u'} {w'} \rangle= \pm \exp{\left(-\int_0^{{z}} \frac{dz'}{(\langle {u} \rangle\mp 1)^4}\right)}.
    \label{eq:Reynolds}
\end{equation}
The control parameter is defined as:
\begin{equation}
    \Lambda_1 = \frac{\widetilde{\nu}\, \widetilde{c}}{\widetilde{d}\,\widetilde{L}}.% \quad \alpha_1=1-\alpha_2, \quad \alpha_2 = \frac{\widetilde{N}^3\left(\widetilde{\nu}+\widetilde{\kappa}\right) \widetilde{d}}{\widetilde{k}\,\widetilde{c^4}},
    %\quad \Lambda_2=\frac{\widetilde{\gamma}\, \widetilde{c}\,\widetilde{d}}{\widetilde{L}},
    \label{eq:paramPlumb}
\end{equation}
%\stepcounter{equation} % Step first to reserve a new number
%\begin{equation}
%    \Lambda_1 = \frac{\nu c}{dL}, \quad \Lambda_2=\frac{\gamma cd}{L}, \quad \alpha_1=1-\alpha_2, \quad \alpha_2 = \frac{N^3\nu d}{kc^4},   
%    \tag{\theequation a--d} \label{eq:paramPlumb}
%\end{equation}
%\stepcounter{equation}
Velocities, lengths and times are measured in units of $\widetilde{c}$, $\widetilde{d}$ and $\widetilde{c}\widetilde{d}/\widetilde{L}$ respectively, where $\widetilde{L}$ is the Reynolds stress contribution from one wave at the bottom of the domain and $\widetilde{d}=\left( \widetilde{N}^3\left(\widetilde{\nu}+\widetilde{\kappa}\right)/\widetilde{k}\widetilde{c^4} \right)^{-1}$ is the damping length of the waves \tck{(e.g. \citet{Plumb1978})}. The latter results from both viscous and thermal dissipation, and its expression holds when diffusive timescales based on the vertical wavenumber are large compared to the wave period. The model we present here differs from that of earlier studies as thermal dissipation is usually neglected in the expression of $\widetilde{d}$. Indeed, as already mentioned in \citet{Couston2018}, the wave-mean flow interaction leading to periodic reversals seems to be favoured for low Prandtl fluids, suggesting different typical temporal dissipative dynamics for the waves ($\widetilde{\nu}$ and $\widetilde{\kappa}$) and the flow ($\widetilde{\nu}$). Some other works also consider linear Rayleigh friction relevant for the atmosphere or laboratory experiments. This introduces another dimensionless parameter $\Lambda_2=\widetilde{\gamma}\, \widetilde{c}\,\widetilde{d}/\widetilde{L}$ where $\widetilde{\gamma}$ is the amplitude of the Rayleigh friction. The Reynolds stress (\ref{eq:Reynolds}) results from the product of perturbations velocities $(u',w')$. These perturbations are taken with respect to the means $(\langle u \rangle,0)$, \tck{so that the total velocity reads $\left(\langle u \rangle +u',w' \right)$.} %It neglects cross wave interaction.

(\ref{eq:Plumb}-\ref{eq:Reynolds}) correspond to the Plumb and McEwan model, notably studied by \citet{Yoden1988, Kim2001, Renaud2019}. Qualitatively, when there is no Rayleigh friction ($\Lambda_2=0$), $\langle u \rangle = 0$ is a solution of (\ref{eq:Plumb}-\ref{eq:Reynolds}) that can become unstable when $\Lambda_1$ is smaller than a threshold $\Lambda_1^c$, i.e. when diffusion is small compared to wave forcing. Introducing Rayleigh friction ($\Lambda_2>0$) tends to stabilise the system and decreases $\Lambda_1^c$. This point has been confirmed by linear stability analysis conducted for various boundary conditions \citep{Semin2018, Renaud2020}. In the vicinity of the threshold, numerical integration of (\ref{eq:Plumb}-\ref{eq:Reynolds}) yields an oscillatory mean flow whose period is of order $1$, which means that in dimensional terms, the oscillation period is of order $\widetilde{c}\widetilde{d}/\widetilde{L}$. As mentioned by \citet{Semin2024}, the fact that the non zero solution past the onset breaks the time translational invariance suggests that the mean velocity undergoes a Hopf bifurcation. They further showed through a weakly nonlinear analysis of (\ref{eq:Plumb}-\ref{eq:Reynolds}) that depending on the ratio of $\Lambda_2/\Lambda_1$ this bifurcation could either be super or subcritical. In the DNS, as we impose a damping layer at the top of the RZ, most of the dynamics is contained between $r_i$ and $r_d$, which is why we consider $\Lambda_2=0$ and expect a supercritical bifurcation. 

This low-dimensional model offers an interesting tool for comparing to our DNS. However, there is one main difference between the two approaches. Convection excites a broad-band, temporally-variable spectrum of waves, opposed to the steady excitation of a single pair of waves assumed here. We will comment on the similarities and differences as we present our results (§ \ref{sec:results}).

\section{Results}\label{sec:results}

\subsection{CZ and wave properties}\label{sec:res:waves}

We will first discuss the convection that develops in our simulations, how it generates waves, and how this impacts the control parameter $\Lambda_1$ introduced above. In the core of the domain, we simulate internally-heated thermal convection. Increasing $Ra$ from $0$ leads the system to bifurcate from the static solution $\mb{u}=\mb{0},T=T_s$ (see Appendix \ref{appA}) to convective patterns and non-zero velocities. As $Ra$ is further increased, turbulence sets in (Figure \ref{fig:1}b).

The typical magnitude of velocity and temperature perturbations can be estimated from simple order of magnitude arguments. Assuming \tck{the velocity to be roughly isotropic ${u_\phi}\sim {u_r} \equiv {u_c}$ with (\ref{eq_DIVU}), with order one length scales in both directions \tck{which we have verified for our simulations}, and that} the dominant balance in the radial component of (\ref{eq_U}) for the CZ is between the inertia and buoyancy terms we obtain 
%\begin{equation}
 %   \frac{u_\phi}{r}\frac{\partial u_r}{\partial \phi} \sim rT \Rightarrow u_\phi u_r \sim T.
 %   \label{eq:res:balanceNS}
%\end{equation}
\begin{equation}
    \frac{{u_\phi}}{{r}}\frac{\partial {u_r}}{\partial \phi} \sim {r}{T} \Rightarrow {u_\phi} {u_r} \sim {T},
    \label{eq:res:balanceNS}
\end{equation}
%Similarly, the heat equation (\ref{eq_THETA}) can be rewritten
%\begin{equation}
 %   \frac{\partial T}{\partial t} +   \nabla \cdot \left( \mb{u}T -\frac{\nabla T}{\left( Ra Pr\right)^{1/2}} \right) = \frac{1}{\left(Ra Pr \right)^{1/2}} \left(1-\frac{T_o}{T_0} \right) I_{r_b} e^{-(r/r_b)^2}. \label{eq:res:balanceT1}
%\end{equation}
%\begin{equation}
%    \frac{\partial \widetilde{T}}{\partial \widetilde{t}} +   \widetilde{\nabla} \cdot \left( \mb{\widetilde{u}}\widetilde{T} -\frac{\widetilde{\nabla} \widetilde{T}}{\left( Ra Pr\right)^{1/2}} \right) = \frac{1}{\left(Ra Pr \right)^{1/2}} \left(1-\frac{T_o}{T_0} \right) I_{r_b} e^{-(r/r_b)^2}. \label{eq:res:balanceT1}
%\end{equation}
%The temperature gradient can be dropped in the left hand side (LHS) for high enough $Ra$, as the temperature profile tends to flatten in this case. This time, it leads to a balance between the first term of the LHS and the RHS. For a statistically stationary case
\noindent which further assumes that the different contributions in the inertia term are of comparable magnitude. Similarly, assuming the dominant balance in the temperature equation (\ref{eq_THETA}) is between advection and internal heating, we find for a statistically stationary case
\begin{equation}
    {\nabla} \cdot \left(\mb{{u}}{T} \right) \sim \left( Ra Pr \right)^{-1/2} \Rightarrow {u_r}{T}\sim \left( Ra Pr\right)^{-1/2}.
    \label{eq:res:balanceT2}
\end{equation}
%\begin{equation}
 %   \nabla \cdot \left(\mb{u}T \right) \sim \left( Ra Pr \right)^{-1/2} \Rightarrow u_rT\sim \left( Ra Pr\right)^{-1/2}.
  %  \label{eq:res:balanceT2}
%\end{equation}
Combining (\ref{eq:res:balanceNS}) and (\ref{eq:res:balanceT2}), we obtain
\begin{align}
    {u_\phi} {u_r} &\sim \left(Ra Pr \right)^{-1/3}, \label{eq:res:CZestimateU} \\
    {T} &\sim \left(Ra Pr \right)^{-1/3}. \label{eq:res:CZestimateT}
\end{align}

%\begin{align}
 %   u_\phi u_r &\sim \left(Ra Pr \right)^{-1/3}, \label{eq:res:CZestimateU} \\
 %   T &\sim \left(Ra Pr \right)^{-1/3}. \label{eq:res:CZestimateT}
%\end{align}

\begin{figure}[t!]
  \centerline{\includegraphics[width=14cm]{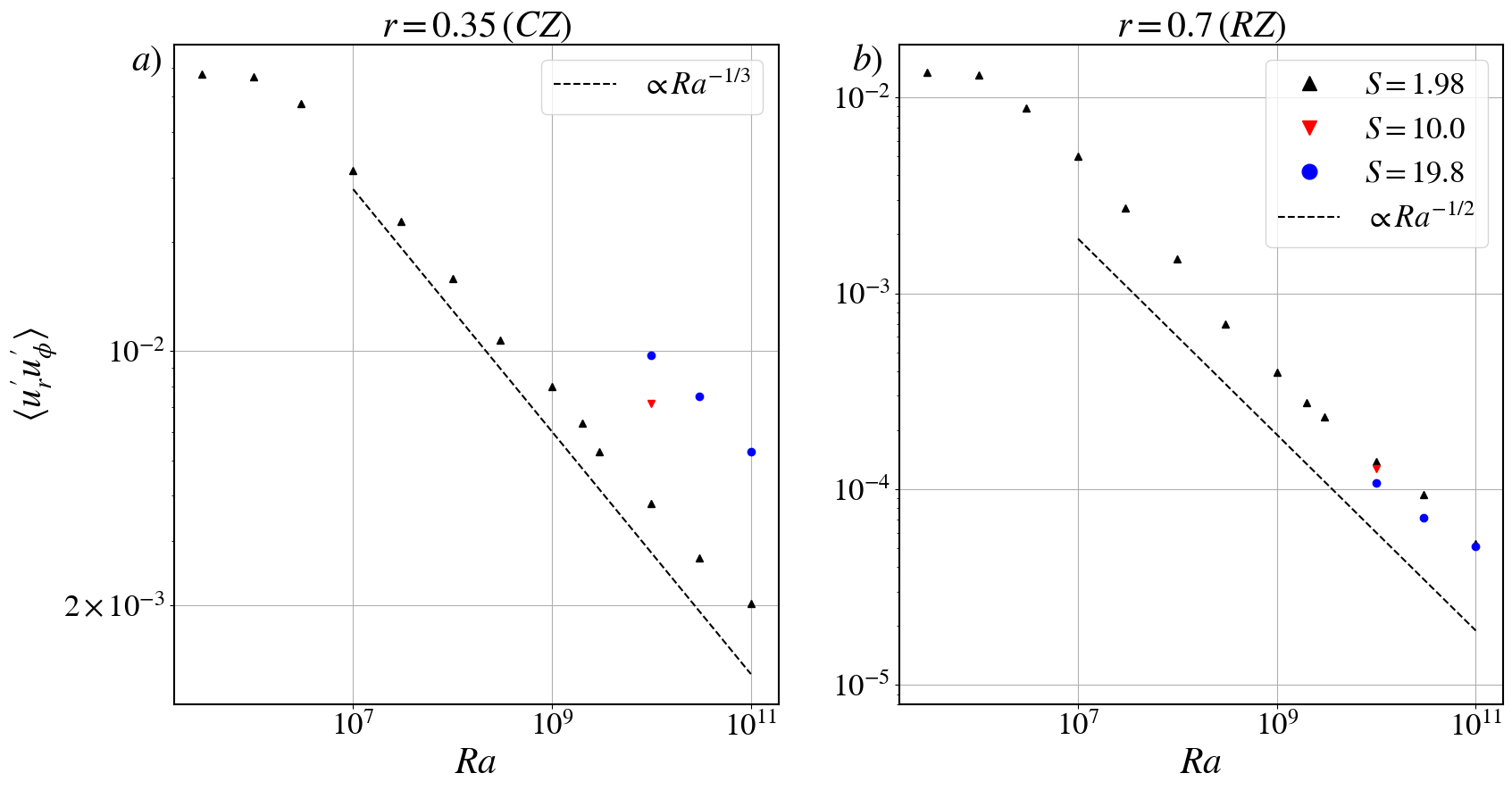}}% Images in 100% size
  \caption{Averaged Reynolds stress for different $Ra$ and $S$, in the CZ (a) and in the RZ (b). Note that contributions from the mean flows have been removed but keeping them gives the same results.} 
\label{fig:3uphiur}
\end{figure}

Note that to derive (\ref{eq:res:CZestimateU}-\ref{eq:res:CZestimateT}) we dropped order one factors. In Figure \ref{fig:3uphiur}a, we plot $\langle {u_\phi'} {u_r'} \rangle$ in the CZ $(r=0.35)$, where $X'$ quantities are deviation from their azimuthal mean $\langle X\rangle$. Note that $\langle u_r u_\phi\rangle = \langle u_r' u_\phi'\rangle$ as $\langle u_r\rangle =0$. The plot shows good agreement with (\ref{eq:res:CZestimateU}). In addition, there is a dependence on $S$, consistent with the earlier study of \citet{Couston2017}. They showed that convective plumes are enhanced for low-stiffness simulations, due to the fact that they can penetrate more easily in the radiative layer as $N$ decreases with $S$ (\ref{eq_N}), consistent with our findings.

The scaling laws (\ref{eq:res:CZestimateU}) and (\ref{eq:res:CZestimateT}) are predictions of the so-called ultimate or ``Gallet'' regime of convection \citep{Kraichnan1962, Spiegel1963,Lepot2018}. The agreement between our simulations and these scaling laws is indeed not surprising as our convective model is similar to previous ones \citep{Bouillaut2019, Hadjerci2024} who demonstrated that for a volumetric heat source, molecular viscosity does not affect the Reynolds number or the heat transfer for flows with high enough $Ra$.
The ultimate regime is expected in problems with no thermal boundary layers \citep{Lepot2018}.
In our polar geometry, there is no bottom boundary at $r=0$, and hence no bottom boundary layer.
%
%In our polar geometry, the confinement of convective motions to the inner layers of the domain means there is no thermal boundary layer at the core of the CZ, crucial to reach this ultimate regime.
%
%While the heat flux is entirely transported by diffusion in the RZ,
Furthermore, the CZ/RZ interface does not act as a thermal boundary layer because there are vertical motions across the interface, and hence the flux is transported by a mix of convection and diffusion.
%
%At the CZ/RZ interface, internal heating precludes diffusion from affecting heat transport.
Note that in our case, the thermal flux equilibrium between the two zones (Appendix \ref{appA}) prevents the spatially-averaged temperature to drift in time, which therefore does not require any cooling \citep{Lepot2018}. We can rewrite the scaling laws for velocity and temperature into the usual ones for the Reynolds number $Re= \widetilde{u_c} \widetilde{r_o}/\widetilde{\nu} \sim \left ( Ra_{eff}/Pr \right)^{1/2}$ and Nusselt number $Nu \sim 1/T \sim \left( Ra_{eff}Pr\right)^{1/2}$, introducing an effective Rayleigh number $Ra_{eff}=Ra {T}$ based on the temperature difference between the centre and the interface. Both scalings are indeed characteristic of the ultimate regime of convection. In other words, the dynamics of the CZ is fundamentally independent of molecular viscosity, which is expected for our setup.

In Figure \ref{fig:3uphiur}b we plot the same quantity $\langle u_r' {u_\phi'} \rangle$ in the stably-stratified region ($r=0.7$). We find a steeper dependency with $Ra$ than in the CZ, with $\langle u_r' {u_\phi'}\rangle \sim Ra^{-1/2}$. It is possible to understand the observed behaviour from the theory of \citet{Lecoanet2013} (see also \citet{Goldreich1990, Couston2018b}). They make a theoretical prediction for the wave energy flux ${F}={u_r} {p}$, which is conserved in the absence of dissipation \citep[e.g.][]{Lighthill1978,LeSaux2023}. Neglecting diffusive effects, they estimate the total wave energy flux to scale as%They indeed predicted the evolution of the wave flux $\widetilde{F}=\widetilde{u_r} \widetilde{p}$. The wave flux is the quantity that is conserved and transported in the inviscid case. It has been discussed many times in the literature \citep{Rogers2013,LeSaux2023}. For the largest eddy of the CZ it reads:

\begin{equation}
    {F} \sim {u_c}^3 \frac{\omega_c}{N},
    \label{eq:F}
\end{equation}

\noindent with ${u_c} \sim \left(RaPr\right)^{-1/6}$ (\ref{eq:res:CZestimateU}) the typical velocity of the CZ and $\omega_c$ the dominant frequency. We will now show this prediction is consistent with the power-law observed in Figure \ref{fig:3uphiur}b. To do so, the pressure must be related to the azimuthal velocity, which is done by taking the horizontal derivative of the linearised horizontal component of (\ref{eq_U}) in the bulk of the RZ (no damping layer) and in the inviscid limit. Assuming that the most significant contribution comes from frequency $\omega =\omega_c$ and horizontal wavenumbers $m=1$ leads to $p=r\omega u_\phi/m$, which indeed gives 
\begin{equation}
    {u_r}{u_\phi}\sim \left( Ra Pr\right)^{-1/2}.
    \label{eq:res:RZestimateU}
\end{equation}
\noindent Thus, similarly to the CZ, the properties of the largest-amplitude waves, which dominate the total wave energy flux and total angular momentum flux, can be explained by diffusion-free arguments. Note this does not necessarily apply to the entire spectrum of IGW, as we discuss next. %However, these arguments show $\Lambda_1$ decreases with $Ra$ (Figure \ref{fig:3uphiur}b), making it more likely to find mean flow reversals at higher $Ra$ (§ \ref{sec:theory}). \\

\begin{figure}[t]
  \centerline{\includegraphics[width=14cm]{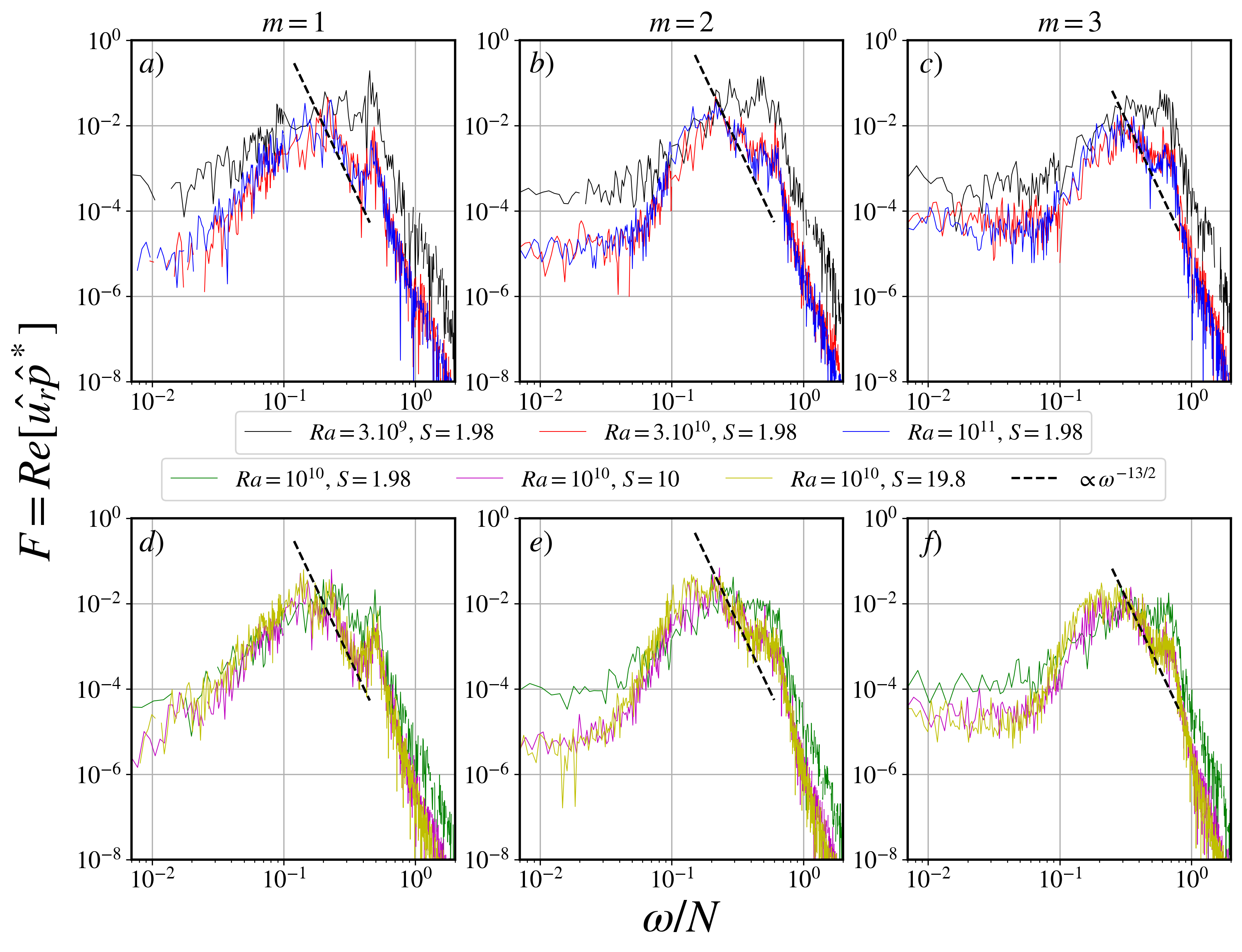}}% Images in 100% size
  \caption{Wave energy flux temporal spectra measurements for varying $Ra$, fixed $S$ (a-c) and fixed $Ra$, varying $S$ (d-f). Spectra are computed in the RZ $(r=0.7)$, and displayed for 3 values of horizontal wavenumbers. The dashed lines show comparisons with predictions by \citet{Lecoanet2013}.}
\label{fig:2spc}
\end{figure}

The dynamics of the waves generated in the CZ and propagating in the RZ can be further characterised by studying spectra. In Figure \ref{fig:2spc} we show frequency spectra of the wave energy flux for simulations of varying $Ra$ (top row) and varying $S$ (bottom row). Reynolds stress or kinetic energy spectra display similar behaviours. Spectra are computed by conducting a double Fourier transform on ${u_r}$ and ${p}$ both along the azimuthal and time coordinates, at $r=0.7$, over one thermal diffusive time. To eliminate contamination from secular variation, we applied a temporal Hann window function. $m=1,2,3$ are the first 3 largest horizontal wavenumbers.

Focusing on the first row, we observe that the different curves tend to collapse onto one another as the Rayleigh number is increased (as ${F}$ is not a positive quantity, we only display positive values). This behaviour is consistent with \citet{Anders2023}: the shape of the spectra, which is the exciting mechanism for the reversing mean flow, converges for high-enough $Ra$. %The fact that we only observe such a convergence for the higher $Ra$ considered suggests that we are in an intermediate regime concerning the dynamics of the RZ, as discussed earlier. 

The second row of Figure \ref{fig:2spc} shows that varying $S$ shifts the spectra to peak at lower frequencies. This can be understood qualitatively by recalling that an increased value of $S$ corresponds to a higher value of $N$ (§ \ref{sec:method}), and hence to a larger separation between the convective frequency $\omega_c$ and that of buoyancy $N$, as the $x$-axis is normalised by $N$. Indeed, we observe that the typical width of the spectra for the highest value of $S$ considered ($S=19.8$, yellow curve) decreases compared to the lower case ($S=1.98$, green curve), leading to spectra being more peaked. Note that the predictions of \citet{Lecoanet2013} assume $\omega_c \ll N$, whereas here we have $N/\omega_c$ at most $3.16$ (for $S=19.8$). Indeed, we find their predictions for the frequency spectra (dashed lines) are at best only valid for a small range of frequencies. \citet{Lecoanet2013} also assume three-dimensional turbulence, though \citet{Lecoanet2021} find similar spectra in two-dimensional Cartesian simulations. The other important point is that their prediction is derived in the inviscid case. Finally, there are peaks slightly below $\omega = N$ corresponding to standing modes that have not been completely removed by the damping layer \citep{Lecoanet2021,Anders2023}.

\begin{figure}[t]
  \centerline{\includegraphics[width=12cm]{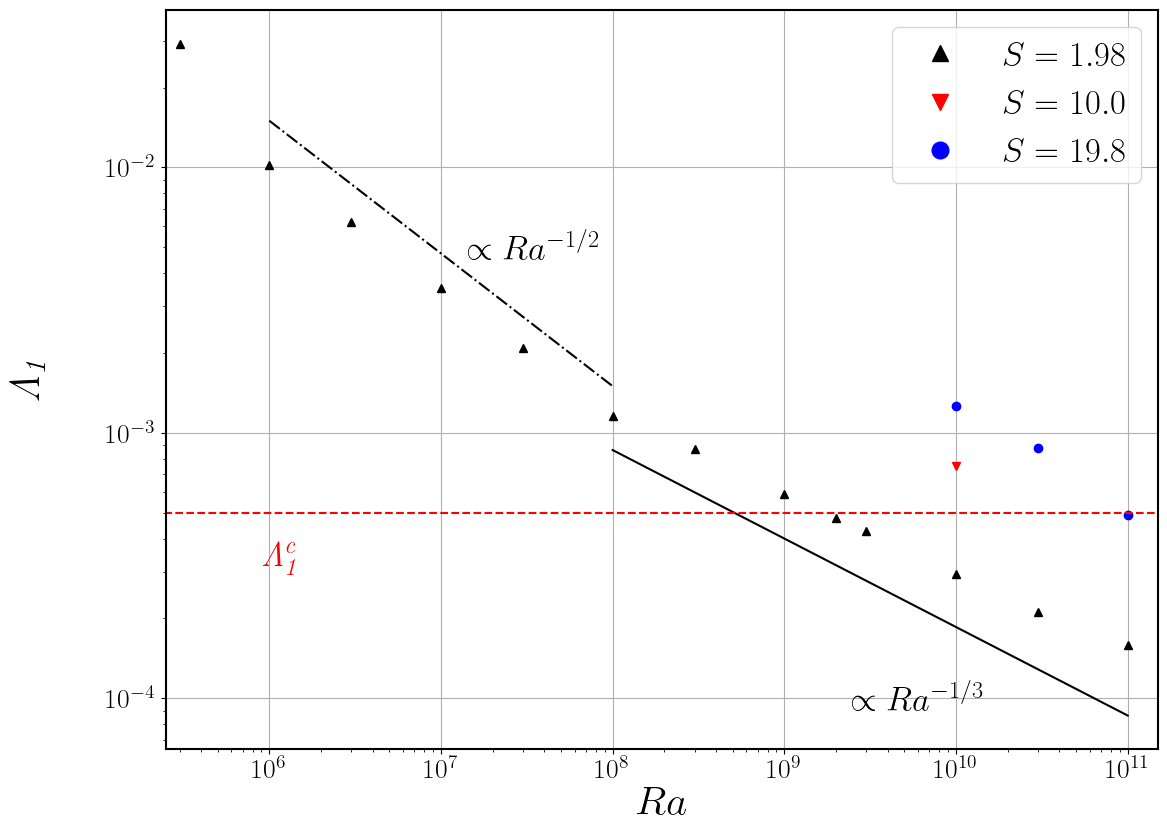}}% Images in 100% size
  \caption{Estimate of $\Lambda_1$ (\ref{eq:paramPlumb}) for the same points as in Figure \ref{fig:3uphiur}, showing how increasing $Ra$ tends to favour low $\Lambda_1$ values and therefore mean flow reversals when $\Lambda_1<\Lambda_1^c=5.10^{-4}$(see (\ref{eq:lbda1DNS}) for details).}
\label{fig:3L1}
\end{figure}

Using these results we can estimate the value of $\Lambda_1$ in our simulations. Rewriting (\ref{eq:paramPlumb}) using the dimensionless parameters of the DNS gives
\begin{equation}
    \Lambda_1 = \frac{1+Pr}{Ra} \frac{1}{\langle {u_r'}{u_\phi'}\rangle}\left(\frac{N}{\overline{\omega}}\right)^3\overline{k}^2.
    \label{eq:lbda1DNS}
\end{equation}
%Note that to derive the previous expression, we assumed $\widetilde{\gamma}=0$ and substituted $\widetilde{\nu}$ into $\widetilde{\nu} + \widetilde{\kappa}$ in the definition of $\widetilde{d}$. This is motivated by a comment in the original model of \citet{Plumb1978} stating that "Generally [...], $\widetilde{\nu}$ is the sum of viscosity and diffusivity".It is in agreement with what we discussed earlier where waves – here through their damping length – are still affected by both Laplacians. Equation (\ref{eq:lbda1DNS}) can be interpreted straightforwardly with what was presented before. 
%We have that $\Lambda_1$ depends on three simulation outputs, $\langle{u_r'}{u_\phi'}\rangle$, $N/\overline{\omega}$ and $\overline{k}$. $\overline{X}$ denotes weighted averages (see below). 
We have that $\Lambda_1$ depends on three simulation outputs, the angular momentum flux $\langle{u_r'}{u_\phi'}\rangle$, the frequency ratio $N/\overline{\omega}$ and the average azimuthal wavenumber $\overline{k}$. $\overline{X}$ denotes weighted averages (see below). 
In Figure \ref{fig:3uphiur} and (\ref{eq:res:RZestimateU}) we show the strong dependence of $\langle {u_r'}{u_\phi'}\rangle$ on $Ra$ and $Pr$. To first order, the ratio $N/\overline{\omega}$ is determined by $S$, and $\overline{k}$ corresponds to the dominant mode of the CZ, which is $m=1$ in our simulations. Note that this reasoning holds for the maxima of spectra, which can be quite dispersed. Here, we rather measured the typical frequencies and azimuthal wavenumbers by conducting weighted averages of the spectra presented in Figure \ref{fig:2spc} in the corresponding direction (see Appendix \ref{appB}), namely $\overline{\omega}\equiv \langle |\omega F |\rangle / \langle | F |\rangle$ and $\overline{k}\equiv \langle |m F |\rangle / (\langle | F |\rangle /0.7)$---we recall that $F$ is measured at $r=0.7$. It enables to obtain a more useful measure of $\Lambda_1$.
%In light of § \ref{sec:theory} and using (\ref{eq:res:RZestimateU}), keeping $S$ and $Pr$ fixed while increasing $Ra$ means that $\Lambda_1 \propto {Ra}^{-1/2}$ (with (\ref{eq:res:RZestimateU})) tends to decrease.
Using (\ref{eq:res:RZestimateU}), we predict that $\Lambda_1$ decreases like ${Ra}^{-1/2}$ at fixed $S$ and $Pr$.
While we recover this scaling for low $Ra$, at higher $Ra$, we find $\Lambda \propto Ra^{-1/3}$. As we showed in Figure \ref{fig:3uphiur}b that $\langle u_r' u_\phi' \rangle \propto Ra^{-1/2}$, $\Lambda_1 \propto Ra^{-1/3}$ means that $\overline{k}$ or $\overline{\omega}$ depend on $Ra$ (see Table \ref{tab:data}). This may be explained by the way we measure $\overline{k}$ and $\overline{\omega}$. Nevertheless, increasing $Ra$ decreases $\Lambda_1$ until it becomes smaller than $\Lambda_1^c $. We estimate $\Lambda_1^c\approx 5.10^{-4}$ in the simulations (red line of Figure \ref{fig:3L1}). The system is then beyond the onset of a reversing mean flow, i.e. the Rayleigh number is higher than a critical Rayleigh number. Higher values of $S$, while increasing the corresponding value of critical Rayleigh numbers, does not affect the picture. Note that (\ref{eq:lbda1DNS}) gives some weight to the interpretation of \citet{Couston2018} who observed favoured mean flows for low Prandtl simulations, as $\Lambda_1 \propto \sqrt{Pr}(1+Pr)$. \tck{On the contrary, laboratory experiments conducted by \citet{Semin2018} led to mean flow reversals using salt stratification, for which the value of the equivalent quantity---the Schmidt number---is rather around $700$. This suggests the quantity which determines the properties of the mean-flow evolution is $\Lambda_1$ rather than $Pr$. \citet{Semin2018} showed that their results could be explained by varying $\Lambda_1$, and we find similar results in our simulations.} 

\subsection{Reversing mean flows}\label{sec:res:meanflow}

\begin{figure}[t]
  \centerline{\includegraphics[width=14cm]{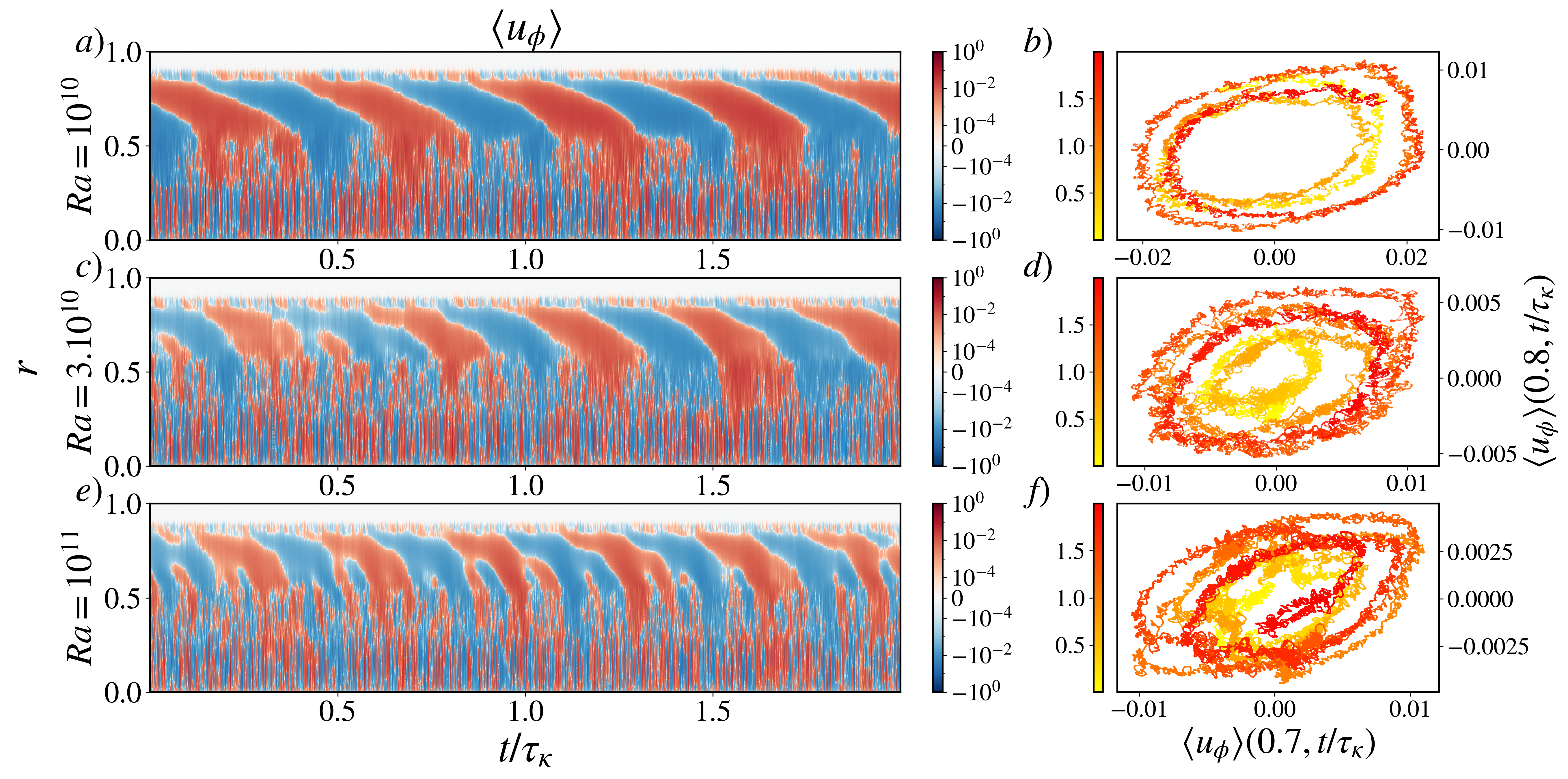}}% Images in 100% size
  \caption{Mean flow visualisation for $S=1.98$. Hovmöller diagrams (Left, a-c-e, colours correspond to flow amplitudes) and phase portraits of local probes of the zonal velocity in the RZ (Right, b-d-f, colours correspond to time) for $Ra=10^{10}$ (a-b), $Ra=3.10^{10}$ (c-d) and $Ra=10^{11}$ (e-f).}
\label{fig:5}
\end{figure}

In Figure \ref{fig:5}, we visualise the mean flows generated in simulations with $S=1.98$ and $Ra\in \{10^{10},3.10^{10},10^{11}\}$. These correspond to simulations where $\Lambda_1$ has become small enough for a mean flow to develop, in agreement with the discussion of § \ref{sec:res:waves} and Figure \ref{fig:3L1}.

In the left column (a-c-e), we plot Hovmöller diagrams showing the zonally averaged azimuthal velocity $\langle{u_\phi}\rangle$ both as a function of space and time. This is a classical type of representation of the problem (see for instance \citet{Wedi2006}), which illustrates how the pattern of the flow evolves. Focusing on panel a, we can describe the velocity evolution. At $t=0$ \tck{(note that this simulation was initialised with the final state of the $Ra=3.10^9$ simulation)}, the mean flow is negative at lower radii in the RZ, and positive at higher radii in the RZ.
We will next discuss the propagation and angular momentum transport of first the prograde (positive phase velocity) waves, and then the retrograde (negative phase velocity) waves.
%The IGW with positive phase velocity propagate from the CZ/RZ interface towards the top of the layer, where they are absorbed leading to a local acceleration of the flow.
%Meanwhile, the waves with negative phase velocity are absorbed lower as they encounter the negative mean flow.
Prograde IGW propagate from the CZ/RZ interface towards the top of the layer. They are absorbed by the medium at a relatively high radius where their absorption leads to a local acceleration of the flow. As time goes by (until $t\approx 0.5 \tau_\kappa$), the absorption process occurs lower as the waves are preferentially absorbed when $\langle u_\phi \rangle \sim c$ (see for instance (\ref{eq:Reynolds})). This leads to the mean flow to propagate downward and even penetrate into the CZ. While these prograde waves are subsequently absorbed at lower radii, retrograde waves are not filtered by the positive mean flow and are able to propagate in the RZ to deposit their angular momentum, with a blue patch that starts to appear at high radii at $t\approx 0.2 \tau_\kappa$. It is the alternation between the absorption processes of prograde and retrograde waves that gives birth to the observed large scale oscillation which repeats afterwards. It is in good agreement with reduced models, the main difference here is that the spectrum of waves is continuous (Figure \ref{fig:2spc}). Note that the typical period of mean flow reversals is comparable to the thermal diffusion time, i.e., much longer than the convective turnover time (see the rapid variations in the CZ). Similar results were obtained by \cite{Couston2018} in Cartesian geometry. This phenomenon is also observed on panels c and e, but the pattern described before is different. Indeed, for the highest $Ra$ reported (panel e), the mean flow has a modified shape, which we interpret as the transition towards a quasi-periodic behaviour. Shorter reversals of the flows can be seen close to the CZ/RZ interface. This is expected in the Plumb-McEwan model \citep{Kim2001,Renaud2019} when $\Lambda_1$ is sufficiently reduced, but here we report a clear example with a self-consistent wave spectrum. \tck{Note that \citet{Chartrand2024} attribute this transition to a quasi-periodic regime by the emergence of new unstable modes in the linearised problem of § \ref{sec:theory}. The comparison is nevertheless complicated by the stochastic nature of the current problem and is left for future work.}

\begin{figure}[t]
  \centerline{\includegraphics[width=14cm]{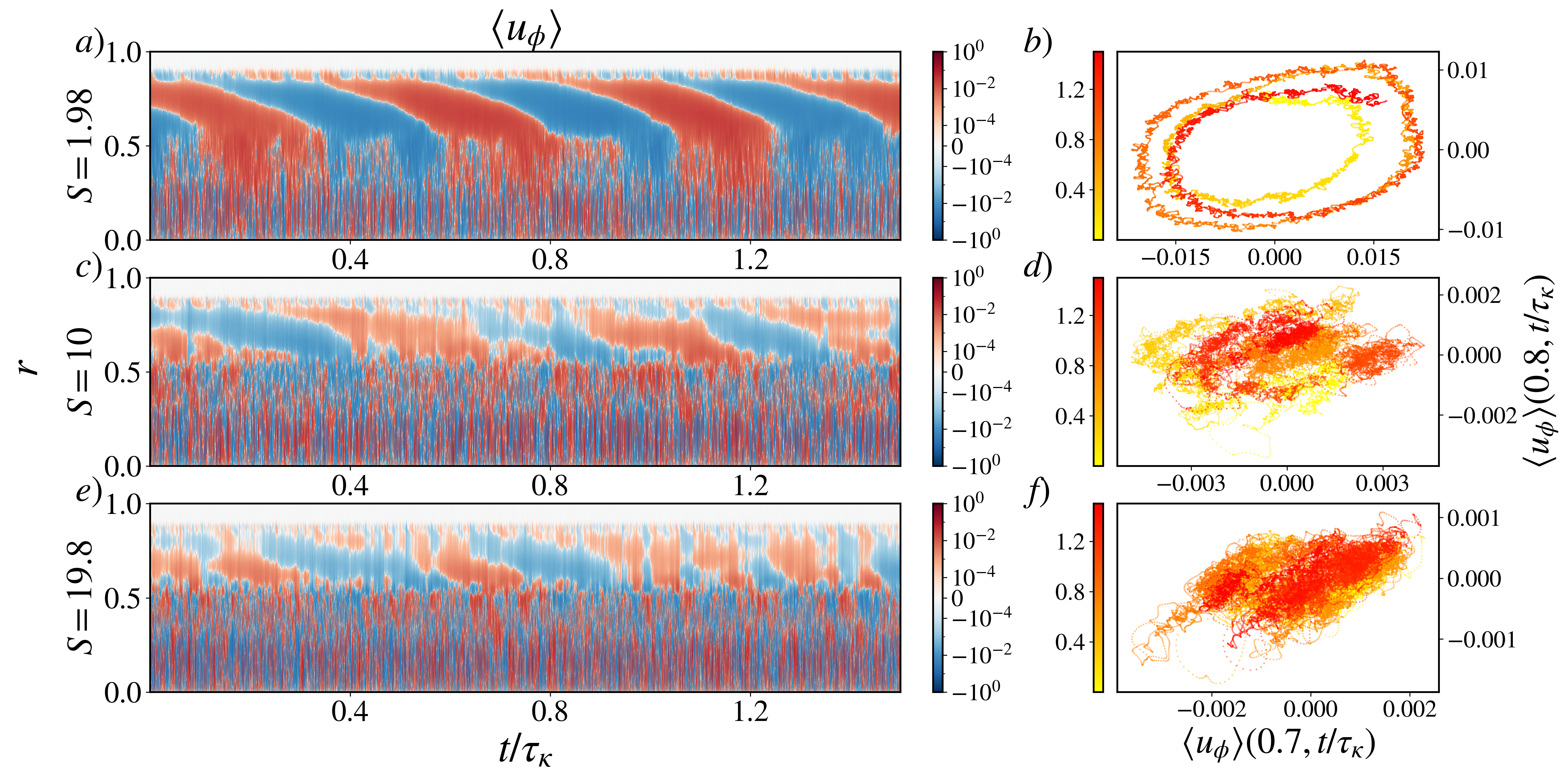}}% Images in 100% size
  \caption{Same as in Figure \ref{fig:5} but for $Ra=10^{10}$, $S=1.98$ (a-b), $S=10$ (c-d) and $S=19.8$ (e-f).}
\label{fig:4}
\end{figure}

In the right column of Figure \ref{fig:5} (panels b-d-f), we plot phase portraits of two local probes of the flow at $r=0.7$ and $r=0.8$. Phase portraits show the emergence of limit cycles and their regularity \citep{Kim2001}. Combining these two representations illustrates how increasing $Ra$ affects the evolution of the flow: the shape shown in panel b displays a regular limit cycle, in agreement with the Hovmöller diagram shown in panel a. There is a single orbit as the flow is periodic. Note the short time scale fluctuations clearly visible in this representation, attributable to fluctuations due to the CZ. The orbit is however modified in panel d and f as $Ra$ is increased, with limit cycles exhibiting now a more complex shape. In panel f, the phase portrait is expected to fill in a torus as the flow is quasi-periodic. Thus, increasing $Ra$ maintains reversals while modifying the period of the flow.

In Figure \ref{fig:4}, we visualise the mean flows generated in simulations with $Ra=10^{10}$ and $S \in \{1.98,10,19.8\}$. While, as shown above, we find regular mean-flow oscillations for low $S$ (panels $a$ and $b$), increasing $S$ causes the mean flow evolution to become increasingly irregular (panels $c$-$f$). That said, the mean flow still exhibits downward-propagating patterns for higher values of $S$. \tck{We expect simulations with $S=10$ or $S=19.8$ would exhibit regular mean flows similar to the $S=1.98$ simulations if they would run with higher $Ra$ corresponding to $\Lambda_1$ sufficiently below $\Lambda_1^c$ (see Figure \ref{fig:3L1})}. Note that in Figure \ref{fig:5} we plot the mean flow evolution for $2$ thermal times, whereas in Figure \ref{fig:4} we plot the mean flow evolution for $1.5$ thermal times. This is in part due to computational constraints, as higher $S$ simulations require higher vertical resolution.

\begin{figure}[t]
  \centerline{\includegraphics[width=14cm]{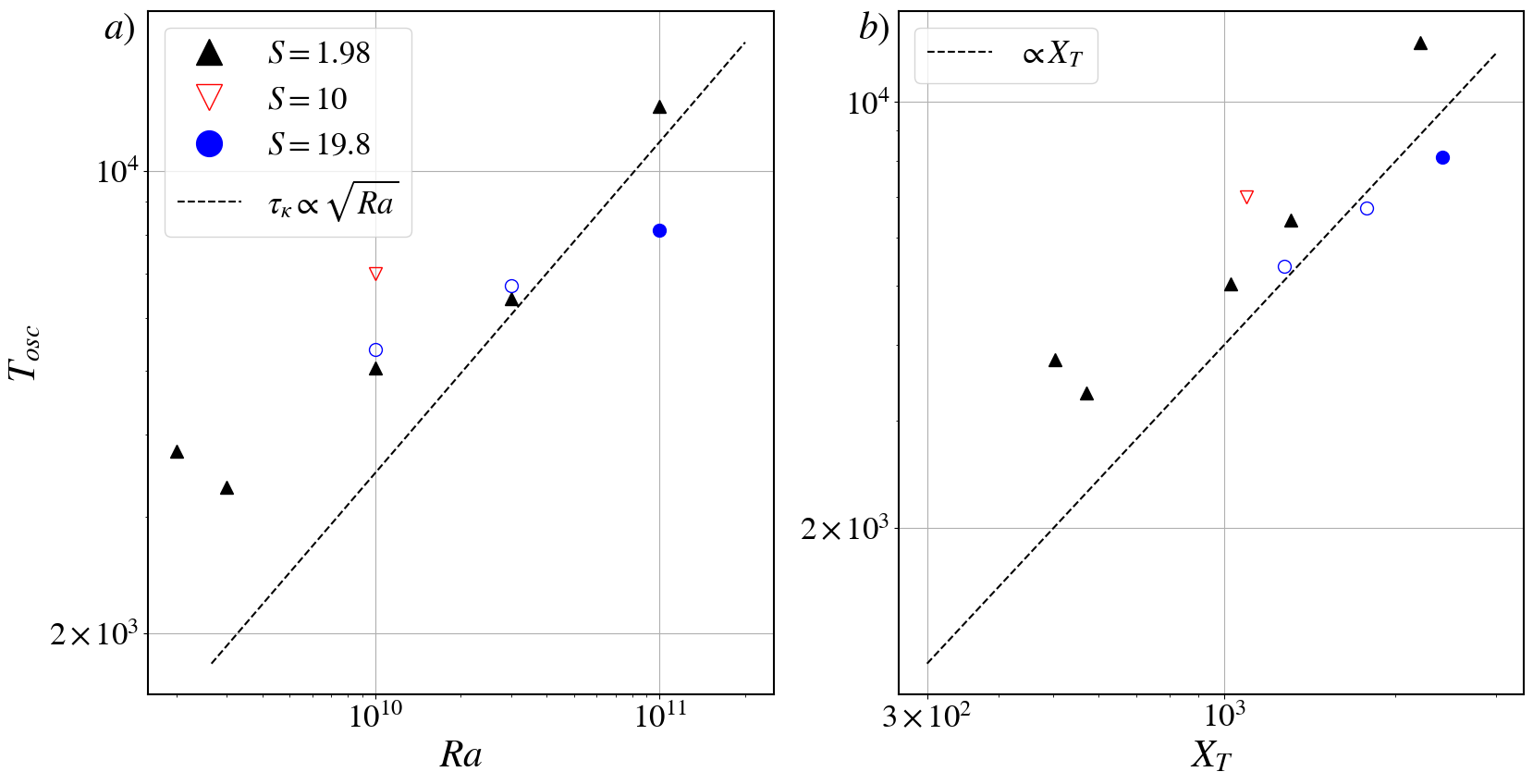}}% Images in 100% size
  \caption{Dominant period evolution. \textit{a)} Measurement in our data set via Fourier transform (see text for details) as a function of $Ra$ for several $S$. \textit{b)} Same as a) but as a function of $X_T$ (\ref{eq:TDNS}). Simulations for which $\Lambda_1$ is higher than $\Lambda_1^c$ are reported as empty symbols.}
\label{fig:6}
\end{figure}

Next we turn our attention to the period of the mean-flow oscillations. We measure the period $T_{osc}$ by finding the frequency of the highest-amplitude peak of the Fourier transform of $u_\phi(r=0.65)$. We find that $T_{osc}$ is insensitive to the choice of radius. We plot $T_{osc}$ for each of our simulations in Figure \ref{fig:6}a. Simulations without coherent mean-flow oscillations ($\Lambda_1>\Lambda_1^c$) correspond to cases where a regular mean flow is difficult to observe, but nevertheless exhibit some activities in their RZ. We also plot with a dashed line $T_{osc} \propto \tau_{\kappa}=\sqrt{RaPr}$, following the observations of Figure \ref{fig:5} where the two time scales appear to be connected.
%It is possible to illustrate this point once again by going back to the reduced model predictions.
%Indeed, assuming that $\widetilde{T_{osc}} \sim \widetilde{c}\widetilde{d}/\widetilde{L}$, one can transcribe this point in the units of the DNS:
%
Indeed, expressing $\widetilde{T_{osc}} \sim \widetilde{c}\widetilde{d}/\widetilde{L}$ in the DNS non-dimensionalisation leads to
\begin{equation}
    %\frac{T_{osc}}{t_0} \sim \left( \frac{\overline{\omega}}{N} \right)^5\frac{\left( Nt_0\right)^2}{\left(\overline{kr_o} \right)^4}\frac{\sqrt{RaPr}}{1+Pr}\frac{1}{\langle\widetilde{u_r}\widetilde{u_\phi}\rangle}\equiv X_T.
    T_{osc} \sim \left( \frac{\overline{\omega}}{N} \right)^5\frac{N^2}{\overline{k} ^4}\frac{\sqrt{RaPr}}{1+Pr}\frac{1}{\langle{u_r'}{u_\phi'}\rangle}\equiv X_T.
    \label{eq:TDNS}
\end{equation}

In Figure \ref{fig:6}b we plot $T_{osc}$ as a function of $X_T$, showing that the simulations agree well with (\ref{eq:TDNS}). 
The combination of the dependencies of both $\langle u_r' u_\phi' \rangle$ and $\tau_\kappa$ on $Ra$ and $Pr$ leads to $T_{osc}\propto (RaPr)^1$.
%While $\langle {u_r'} {u_\phi'} \rangle$ affects the oscillation timescale, the dimensionless thermal diffusive time interestingly appears in the RHS of (\ref{eq:TDNS}), as $\sqrt{RaPr}$.
The diffusive time scale constrains the oscillation of the mean flow through the typical damping length of the waves, which for low $Pr$ fluids is dominated by thermal diffusion.
(\ref{eq:TDNS}) was derived in the framework of the Hopf bifurcation which means that in the vicinity of the threshold, ${T_{osc}^{1D}}\sim O(1)$ (we remind that the 1D and DNS have different units). \tck{Besides showing the good agreement with earlier reduced models, our DNS allows to compute the prefactor in (\ref{eq:TDNS}), $T_{osc}=4X_T$.}
%The combination of the dependencies of both $\langle u_r' u_\phi' \rangle$ and $\tau_\kappa$ on $Ra$ and $Pr$ leads $T_{osc}\propto (RaPr)^\alpha$. At dominant order, our data suggests $\alpha \approx 1$.
%Consequently, at leading order, $T_{osc}\propto RaPr$.

%Interestingly, the dimensionless thermal diffusive time does appear in the RHS of (\ref{eq:TDNS}), as $\sqrt{RaPr}$, explaining how it affects the oscillation of the mean flow, through the typical damping length of the waves, which for low $Pr$ fluids is dominated by thermal diffusion.

\begin{figure}[t]
  \centerline{\includegraphics[width=14cm]{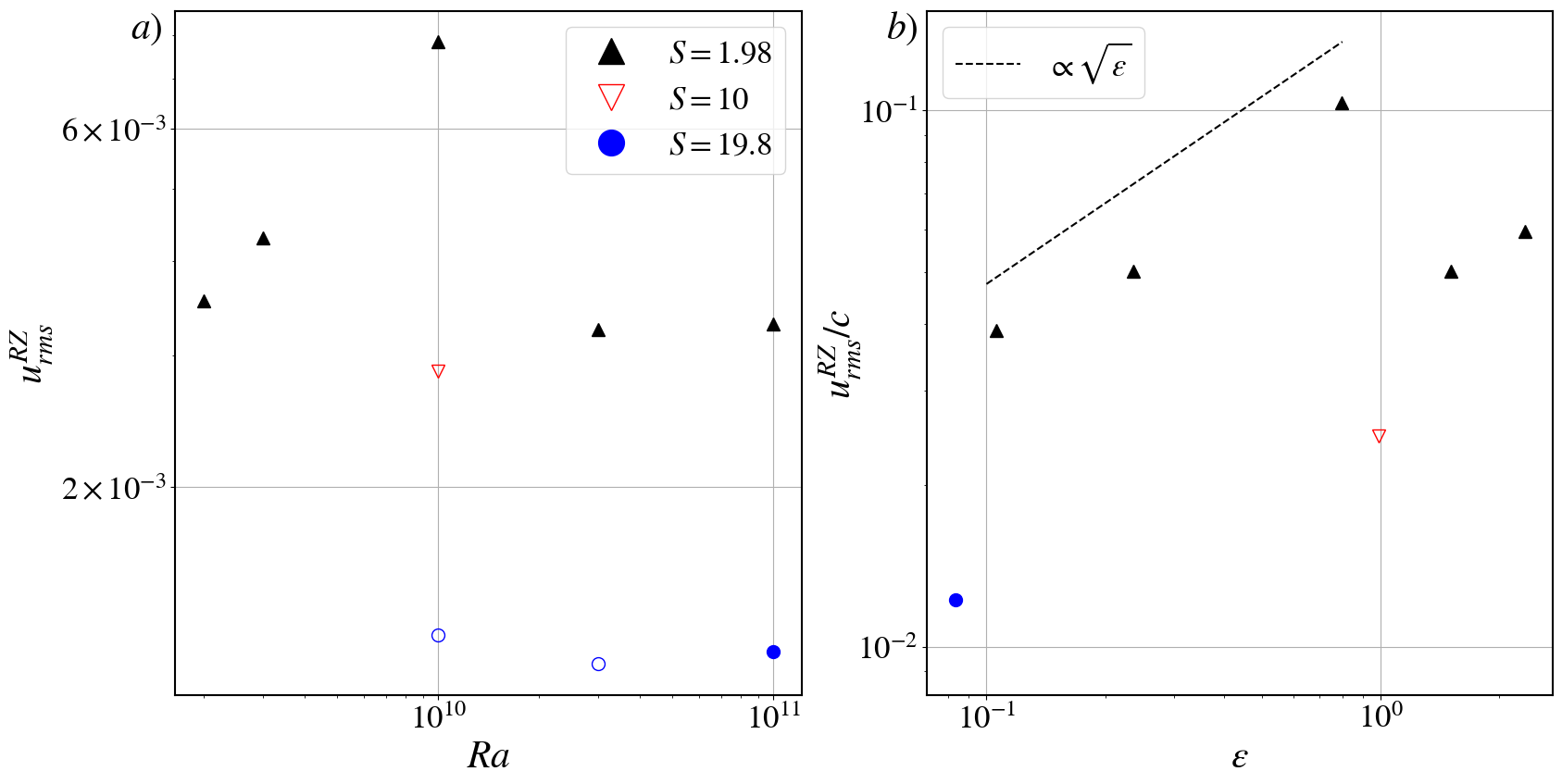}}% Images in 100% size
  \caption{Mean flow velocity evolution. \textit{a)} Root mean square of the RZ zonal velocity (measured from $r=0.6$ to $r=1$ in our data set) as a function of $Ra$ for several $S$. \textit{b)} $u_{rms}^{RZ}/c={u}_{rms}^{RZ}/\left(\overline{\omega}/\overline{k} \right)$ (see text for details) against the distance from the onset $\varepsilon=(\Lambda_1^c-\Lambda_1)/\Lambda_1$.}
\label{fig:7}
\end{figure}

The Plumb-McEwan model also makes predictions for the velocity amplitude of the mean flow. Figure \ref{fig:7}a shows the rms zonal velocity $\langle {u_\phi}\rangle$ in the RZ as a function of $Ra$ for several $S$.
%As for the period, we show the result in Figure \ref{fig:7}a as a function of $Ra$ for several $S$, where we measure the root mean square zonal velocity $\langle {u_\phi}\rangle$ in the RZ.
On the one hand, for $S=1.98$, the flow increases in amplitude until $Ra=10^{10}$. We can interpret this in terms of the 1D model for which the control parameter is $1/\Lambda_1$.
Thus, for this Hopf bifurcation, we expect the velocity to increase as the square root of the supercriticality,
%The fact that it corresponds to a Hopf bifurcation states that past the onset, the velocity should increase as the square root of the distance to the threshold,
i.e. $u^{RZ}_{rms}/c \propto \sqrt{(\Lambda_1^c-\Lambda_1)/\Lambda_1}\equiv \sqrt{\varepsilon}$. We show in panel b that this is consistent with the three lowest $Ra$ at $S=1.98$, though this result is very sensitive to the precise value of $\Lambda_1^c$. However, for larger $Ra>10^{10}$ we find smaller amplitude mean flows, which we attribute to the emergence of new patterns and the transition towards a quasi-periodic state. \tck{Note that what appears to be a secondary transition between $Ra=10^{10}$ and $Ra=3.10^{10}$ could be due to the fact that the wave damping length starts to be comparable to the size of the RZ, as it reads $\widetilde{d}/(\widetilde{r_o}-\widetilde{r_i}) = \left(\overline{\omega}/N\right)^4 N/\overline{k}^3\sqrt{RaPr}/(1+Pr)/(1-r_i)$. \citet{Chartrand2024} indeed showed that more complex bifurcations could occur when it is the case, which could account for the non-monotonic velocity variations with $Ra$ in the present work.} On the other hand, increasing $S$ seems to reduce the amplitude of the flow through a higher value of the Brunt-Väisälä frequency. 

We want to emphasise one the striking finding of our paper that, DNS, although involving a continuous spectrum of IGW due to the CZ, are well captured by the Plumb and McEwan model. It hints that spectra, through their dominant frequency, act as an effective monochromatic forcing. This point should however be nuanced and remains qualitative, particularly for the velocity. The ability of the reduced model to quantitatively compare to global simulations outputs was tested by \citet{Couston2018} who extracted DNS spectra of various frequencies and horizontal wavenumbers. Concretely, it consists in integrating numerically the model of § \ref{sec:theory} while summing contributions of the Reynolds stress (\ref{eq:Reynolds}) from several waves, taking for the prefactors the values measured in DNS. While they were able to reproduce mean flow reversals with this mixed DNS-Plumb and McEwan model, the comparison was not perfect. One possible reasons could be that, as in our case, the important level of fluctuations due to the CZ limits the comparison with the 1D model. Indeed (see § \ref{sec:theory}), the latter considers deterministic dynamics, i.e. neglects any stochastic processes. We posit that the fluctuations of the CZ, for instance acting through temporal variations of $\langle {u_r'} {u_\phi'}\rangle$, affects the classical picture. Their effect could be thought of as a multiplicative noise, which we discuss in § \ref{sec:ccl}. The present point is that in the perspective of understanding the system as resulting from a Hopf bifurcation, the notion of a threshold only makes sense statistically. This may explain why the run at $S=10$, $Ra=10^{10}$ has $\Lambda_1>\Lambda_1^c$ (see Appendix \ref{appB}), i.e. should be classified as stable, but nevertheless exhibits features (see again Figure \ref{fig:4}c) that are typical of a reversing mean flow. Fluctuations are also apparent in the phase portraits of Figures \ref{fig:5} and \ref{fig:4} where noise superimposed to limit cycles is clearly visible. 

\begin{figure}[t]
  \centerline{\includegraphics[width=14cm]{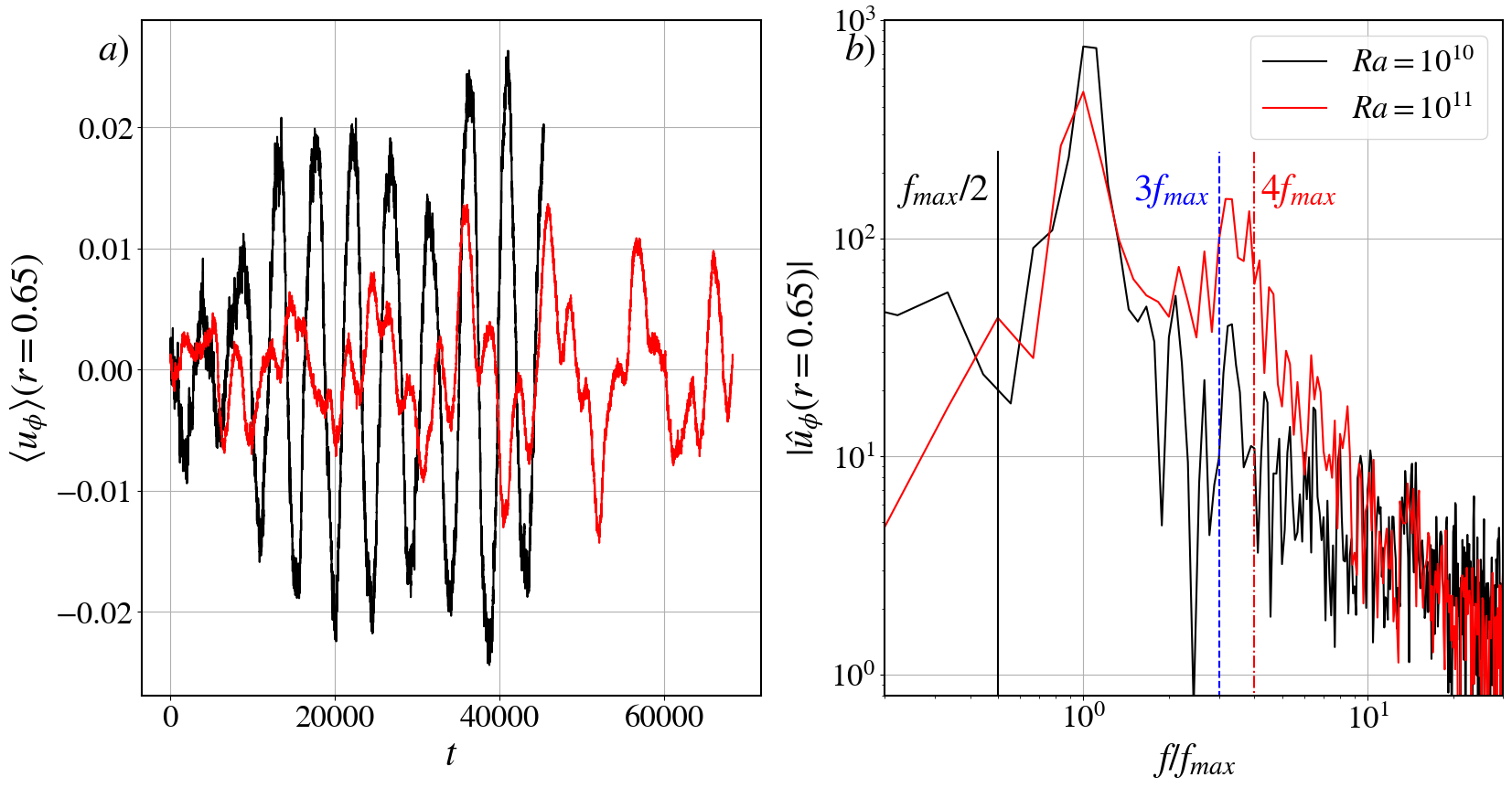}}% Images in 100% size
  \caption{Temporal behaviours of the mean flow. \textit{a)} Local probes of the zonally averaged velocity at $r=0.65$ as a function of time for $Ra=10^{10}$ and $Ra=10^{11}$ both at $S=1.98$. \textit{b)} Corresponding Fourier transform for the signals reported in a). The x-axis is normalised by the value of the dominant frequency in each case.}
\label{fig:8}
\end{figure}

We show in Figure \ref{fig:8} $\langle {u_\phi} \rangle(r=0.65)$ for the simulations $S=1.98$, $Ra=10^{10}$ and $Ra=10^{11}$, corresponding to the Hovmöller diagrams reported in Figure \ref{fig:5}a and e. We use these data to measure the mean-flow oscillation periods. We show the corresponding Fourier transform in Figure \ref{fig:8}b. The time series illustrate the time scale separation which is at the basis of past reduced models. Convective motions produce variations on timescales $O(1)$. It is these rapid variations that lead to the much longer oscillation of $O(10^4)$ corresponding to the reversing mean flow. On the right, the Fourier transforms exhibit two clear peaks. Interestingly, even the peak-to-peak amplitude fluctuates, as the results of the CZ noise discussed earlier. We also see that the typical mean-flow velocity decreases as Ra increases (Figure \ref{fig:7}b). Our interpretation is that the system undergoes a second transition in this case, leading to the emergence of a second noticeable peak in the Fourier transform. Indeed, a higher frequency emerges on the red curve of Figure \ref{fig:7}b, which seems to be situated between $3$ and $4$ $f_{max}$. The fact that it does not correspond to an integer multiple of $f_{max}$ suggests the transition to a quasi-periodic regime, as one would rather expect clear harmonics---integer multiples of $f_{max}$---if the system were to remain periodic. This is compatible with the temporal probe of panel a, which exhibits shorter oscillations between each global extremum, that are absent for $Ra=10^{10}$, as well as with the panel e of Figure \ref{fig:5}, where shorter oscillations appear at $r=0.65$.

We also report a third peak lower than $f_{max}$ for the $Ra=10^{11}$ simulation at $f_{max}/2$, compatible with a \tck{phase-locking} scenario. This was already observed in reduced models for some range of the parameters \citep{Kim2001}, and is a classical scenario transition towards chaotic states. Note however that caution must be considered about this point as very long statistics---numerically demanding---could improve the quality of the signals. It is besides not impossible that fluctuations could lead this $1/2$ peak to be only a transient, or that other quasi-periodic states could emerge, corresponding to different ratios of frequencies \citep{Kim2001,Renaud2019}. While increasing $Ra$ from $10^{10}$ to $10^{11}$ modifies the picture with the onset of different patterns, both visually (Figure \ref{fig:5}) and in the spectra (Figure \ref{fig:8}b), it is not clear yet why the velocity decreases in this case. Further investigation of the emergence of peculiar frequencies could be an interesting perspective in future laboratory experiments which can produce longer datasets than DNS.

\section{Discussion and perspectives}
\label{sec:ccl}

In this paper, we studied the dynamics of mean flow oscillation in a stably stratified layer, through the coupling with a convectively unstable layer. Extending earlier work by \citet{Couston2018} in Cartesian geometry, we worked in polar geometry, and via a parametric survey, investigated the role of varying the strength of convection ($Ra$) and the buoyancy frequency ($S$) on the mean flow oscillations. Our results showed overall good \tck{qualitative} agreement with the weakly nonlinear analysis of a reduced model of the problem conducted by \citet{Semin2018}. The mean flow develops as a Hopf bifurcation, leading to a simple dependence of its period and velocity amplitude in the vicinity of the threshold (Figures \ref{fig:6} and \ref{fig:7}). Further from the bifurcation threshold, the system develops more interesting nonlinear behaviours (Figure \ref{fig:8}). As this rich topic is at the boundary between Geophysical and Astrophysical Fluid Dynamics and nonlinear physics, we now discuss some directions about possible future works.

A broader exploration of the parameter regime of this problem could be conducted. Even if going towards astrophysical values for $Ra$ and $S$ is out of reach---broadly expected to be larger than $10^{20}$ and $10^6$---it would be interesting to run simulations at $S=O(100)$ in the context of mean flows, as the CZ was observed to be barely affected by the RZ in this regime \citep{Couston2017}. Those simulations would nevertheless be very demanding numerically with high spatial resolutions for very long temporal integrations. 

Besides the control parameters that we varied here ($Ra$ and $S$) or that were varied in the past ($Pr$ by \citet{Couston2018}), it is worth commenting the additional ones used in our model. The internal heating profile has a spatial extension controlled by $r_b=0.1$ that was not changed. As discussed in § \ref{sec:method}, $r_b$ value has been chosen so that $r_b<r_i$ to confine heating to the CZ, but varying its value could change things quantitatively. Nevertheless, the fact that mixed layers models can lead to a reversing mean flow with two types of convection---internal heating here or classical Rayleigh Bénard boundary forcing \citep{Couston2018}---suggests that the global picture does not depend on these precise details of the CZ. 

Another important assumption of our model is the form of the damping layer. Even if it depends on two parameters ($r_d$ and $\delta$), the fact that it is confined to the very top of the domain makes it compatible with the classical picture of the QBO. Indeed, the wave absorption mechanism described in the discussion of Figure \ref{fig:5} (see also \citet{Vallis2017}) implies that there is no downwards propagation of the waves, i.e. that the mean flow is not affected by what occurs above the highest level of wave absorption. This description is relevant for the atmosphere where the waves are expected to break if they reach high-enough altitudes. However, in stars, this is not necessarily the case as reflections can occur. Indeed, standing modes are observed via asteroseismology (see for instance \citet{Aerts2021}). Standing modes are still present in our simulations due to the reflection with the top boundary (see Figure \ref{fig:2spc}). A further investigation of this problem could be to compare simulations relevant for planets, with a damping layer, to simulations relevant for stars, with a reflecting boundary. This could be for instance done in the light of recent results by \citet{Chartrand2024} who studied the effect of the ratio of the wave damping length to the size of the domain, which would amount to varying $d/(r_d-r_i)$ in our case. Another interesting possibility could be to examine other techniques to more effectively prevent reflections (e.g. \citet{Berenger1994}).

Different dissipation mechanisms have been studied in different models. The Plumb-McEwan model considers a mix between diffusion and friction which as described above (§ \ref{sec:theory}) can be non-dimensionalised using $\Lambda_1$ and $\Lambda_2$ respectively. For most of the RZ between $r_i$ and $r_d$, our model does not have friction. Thus, in the notation of \citet{Semin2018}, we have $\Lambda_2=0$.
This means the effect of changing $N$ is different in the two systems.
For us, varying $S$, and hence $N$, only modifies $\Lambda_1$ as $\Lambda_2=0$.
Thus, we always find supercritical bifurcations.
In contrast, when \citet{Semin2018} vary $N$ in their laboratory experiments, this changes both $\Lambda_1$ and $\Lambda_2$, and allows for the transition from supercritical to subcritical bifurcations that are not observed in our system.

%In our case, we observe a modification of the onset of the mean flow in terms of $Ra$ while varying $S$---or equivalently $N$---because $\Lambda_1$ (\ref{eq:lbda1DNS}) depends on $N$. This is fundamentally different from \citet{Semin2018} in the sense that varying $N$ in their lab experiment strongly affects both $\Lambda_1$ and $\Lambda_2$, and allows for subcritical behaviours that were not observed here. 

The qualitative agreement we managed to highlight through the present work between the study of \citet{Semin2018} and ours raises the more general question of how to compare complex global simulations involving a large number of modes to simpler reduced models considering only a few. As DNS are very demanding numerically, reduced models offer a complimentary approach. %interesting compromises.
If the Plumb and McEwan approach has already proven to capture the dynamics of the QBO despite its apparent simplicity \citep{Baldwin2001, Renaud2019}, we believe that our model could serve as a starting point for improvement as the CZ is self-consistently simulated. Related to the fluctuant dynamics of the latter and the discussion around Figure \ref{fig:8}, future works should focus on characterising the effect of the noise on the generation of the mean flow. This could be done via a twofold approach. First by modifying the model of § \ref{sec:theory} in order to study the effect of a multiplicative noise (e.g. \citet{Fauve2017}) on the mechanism, as was for instance done by \citet{Ewetola2024}. Note that stochastic dynamics were also considered in \citet{Renaud2018} but through additive noise, showing interesting complex behaviours. Second, DNS could be used to infer the precise properties of the noise induced by the CZ, in the spirit of laboratory experiments focusing on fluctuations \citep{Aumaitre2003} or DNS \citep{Labarre2023}, through probability density functions. Similar characteristics were already measured in \citet{Couston2018} that could be used to set the noise of a stochastic Plumb McEwan model.

To conclude this study, we finally discuss the initial motivation for this work (§ \ref{sec:intro}) which is stellar fluid dynamics and astrophysical applications. Indeed, the geometry of the domain was chosen to model a massive star with a convective core and a radiative envelope. Many simplifications were assumed that could be investigated in the future by adding physical effects to the current model, for instance 3D dynamics, rotation, or even magnetic fields. In this work, we employ the Boussinesq approximation and a piecewise-linear equation of state to model the interaction between a CZ and a RZ. Thus, we assume that density variations are small between the core and the surface. While this is not the case for stars (see for instance \citet{Anders2023}), our approach allows to tackle the coupling between the two zones self-consistently. In a star, the fluid transitions from convective to stably-stratified due to large changes in the background thermodynamic profiles. Going towards fully-compressible, or in an intermediate step anelastic simulations, may therefore appear as an appealing next step for astrophysical considerations. 

\tck{Although here we considered a central CZ surrounded by an outer RZ, there are also natural systems (e.g., solar-type stars) in which a convective envelope lies above a central radiative zone. While adapting the present Boussinesq model would presumably still allow for mean flow reversals---albeit with some differences due to changes in the nature of the CZ---the solar-like and massive stars cases would differ a lot when taking into account density variations. In the former, gravity waves excited by convection propagate downwards, in a direction where density increases, whereas it is the other way around in the latter, similarly to what happens in the Earth's atmosphere. In both stellar cases however, the criterion for gravity waves to reach high amplitudes and break is that the wave luminosity be higher than the radiative luminosity, which is not expected in most stellar environments \citep{Lecoanet2019}}. \tck{In the solar-like case, \citet{Rogers2006} argued that they observed hints of a mean flow reversal in the stably-stratified layer of an anelastic simulation.}
%Using the work by \citet{Plumb1977}, this suggests that observing reversing mean flows in the solar-like case would be harder than in massive stars, as the amplitude of the forcing increases with the radius. Note that this increase could however be counteracted by radiative diffusion \citep{Lecoanet2019}}. \tcb{In the solar-like case, \citet{Rogers2006} argued that they observed hints of a mean flow reversal in the stably-stratified layer of an anelastic simulation.}

While it is not possible to run simulations at realistic parameter values, it is possible to extrapolate our results to astrophysical parameters. Using the characteristics of a 15 solar mass star \citep{Lecoanet2023,Anders2023}, a direct numerical application of (\ref{eq:lbda1DNS}) leads to $\Lambda_1 \approx 2.10^{-6} \ll \Lambda_1^c\tck{\approx 5.10^{-4}}$, \tck{suggesting that the supercriticality of stars is only of order $100$, despite asymptotically large values of $Ra$ and $S$.} Oscillations of the kind reported here are then expected, with a typical period of order $5.10^6$ years (\ref{eq:TDNS}). Note that this estimate is an upper limit of the period and could differ in a real star. For instance, we have discussed that the behaviour of $\Lambda_1$ changes with $Ra$ (Figure \ref{fig:3L1}). In our simulations, waves interact with the mean flow diffusively, whereas in atmospherical and astrophysical applications they are thought to interact with the mean flow via critical layer interactions and wave breaking. Indeed, the fact that $u_{rms}^{RZ}/c \ll 1$ (Figure \ref{fig:7}) combined with the discussion about the oscillation being related to the wave damping length---through the thermal diffusive time (\ref{eq:TDNS})---highlights that waves are absorbed via diffusive processes and not wave breaking. This question will be addressed in a future work by focusing solely on the RZ, where an emphasis will be put on how wave breaking affects the mean flow. It could be interesting to connect breaking and the reduced model approach, through the use of turbulent diffusion coefficients. These could drastically reduce the previous estimate of the period.

In the tradition of the history of wave-induced mean flows, adopting various approaches with different scales, through 1D or global DNS, seems the adequate philosophy to tackle this rich nonlinear problem.

\begin{bmhead}[Funding]
FD is supported by the Simons Foundation grant SFI-MPS-T-MPS-00007353 and BSF grant No. 2022107. DL is partially supported by NSF AAG grant AST-2405812, Sloan Foundation grant FG-2024-21548 and Simons Foundation grant SFI-MPS-T-MPS-00007353.
\end{bmhead}

\begin{bmhead}[Declaration of Interests]
The authors report no conflict of interest.
\end{bmhead}

% \backsection[Supplementary data]{\label{SupMat}Supplementary material and movies are available at \\https://doi.org/10.1017/jfm.2019...}
%
% \backsection[Acknowledgements]{Acknowledgements may be included at the end of the paper, before the References section or any appendices. Several anonymous individuals are thanked for contributions to these instructions.}
%
% \backsection[Funding]{Please provide details of the sources of financial support for all authors, including grant numbers. Where no specific funding has been provided for research, please provide the following statement: "This research received no specific grant from any funding agency, commercial or not-for-profit sectors." }
%
% \backsection[Declaration of interests]{A Competing Interests statement is now mandatory in the manuscript PDF. Please note that if there are no conflicts of interest, the declaration in your PDF should read as follows: {\bf Declaration of Interests}. The authors report no conflict of interest.}
%
% \backsection[Data availability statement]{The data that support the findings of this study are openly available in [repository name] at http://doi.org/[doi], reference number [reference number]. See JFM's \href{https://www.cambridge.org/core/journals/journal-of-fluid-mechanics/information/journal-policies/research-transparency}{research transparency policy} for more information}
%
% \backsection[Author ORCIDs]{Authors may include the ORCID identifers as follows.  F. Smith, https://orcid.org/0000-0001-2345-6789; B. Jones, https://orcid.org/0000-0009-8765-4321}
%
% \backsection[Author contributions]{Authors may include details of the contributions made by each author to the manuscript'}

\begin{appen}

\section{Static temperature profile and interface location}\label{appA}

%\subsection{Temperature unit}

Here we will describe the diffusive temperature equilibrium, and how it is used to define $\widetilde{T_0}$, which we use to non-dimensionalise temperatures in our model.

When the Rayleigh number is below the onset of convection, one can solve for the temperature profile $\widetilde{T_s}$
\begin{equation}
    \widetilde{\Delta}\, \widetilde{T_s} = \frac{1}{\widetilde{r}}\frac{d}{d\,\widetilde{r}}\left(\widetilde{r}\,\frac{d\,\widetilde{T_s}}{d\widetilde{r}}\right)= \frac{-\widetilde{Q}(\widetilde{r})}{\widetilde{\kappa}},
\end{equation}
which straightforwardly integrates to
\begin{equation}
    \widetilde{T_s}(\widetilde{r}) = \widetilde{T_o} + \int_{\widetilde{r}}^{\widetilde{r_o}}\left( \int_0^{\widetilde{r'}} \frac{\widetilde{Q}(\widetilde{r''})\widetilde{r''}d\widetilde{r''}}{\widetilde{\kappa}}\right)\frac{d\widetilde{r'}}{\widetilde{r'}}.
    \label{eq:apdx:Ts}
\end{equation}
Note that we keep $\widetilde{Q}(\widetilde{r})$ as a generic function of $\widetilde{r}$ for generality but that our simulations use $\widetilde{Q}(\widetilde{r})=\widetilde{Q_0}e^{-(\widetilde{r}/\widetilde{r_b})^2}$. We then define $\widetilde{T_0}$ as
\begin{equation}
    \widetilde{T_0} \equiv \widetilde{T_s}(0)-\widetilde{T_s}(\widetilde{r_i^s}) = \frac{1}{1+\frac{\widetilde{T_i}-\widetilde{T_o}}{\widetilde{T_0}}}\int_0^{\widetilde{r_o}}\left( \int_0^{\widetilde{r'}} \frac{\widetilde{Q}(\widetilde{r''})\widetilde{r''}d\widetilde{r''}}{\widetilde{\kappa}}\right)\frac{d\widetilde{r'}}{\widetilde{r'}}.
\end{equation}
\noindent The latter expression can be used to make the system (\ref{eq_U_dim}-\ref{eq_DIVU_dim}) dimensionless, where the following dimensionless integral appears:

\begin{equation}
    I_{r_b} = \frac{1}{\int_0^{1}\left( \int_0^{r'} e^{-(r''/0.1)^2}r''dr''\right)\frac{dr'}{r'}}.
\end{equation}

The previous results can be used to infer the location of the interface between the RZ and the CZ. We assume the flux is carried by convection for $r<r_i$ and is carried by conduction for $r>r_i$. After some algebra, we find

\begin{equation}
	\ln (r_i) \frac{\int_0^{r_i} e^{-(r/0.1)^2} rdr}{\int_0^1 \left( \int_0^r e^{-(r'/0.1)^2}r'dr' \right) dr/r} = \frac{T_o}{1-T_o},
    \label{eq:apdx:ri}
\end{equation}
\noindent an implicit equation for $r_i$. Figure \ref{fig:apdx1} shows very good agreement between our measurement of the CZ/RZ interface position in DNS---tracked through the value $T(r=r_i)=0$---and this analytical prediction. The figure illustrates how $T_o$ controls the geometry of the domain. For $T_o=0$ we obtain a purely convective domain, while the RZ becomes larger when $|T_o|$ increases.  

\clearpage
\begin{figure}
  \centerline{\includegraphics[width=9cm]{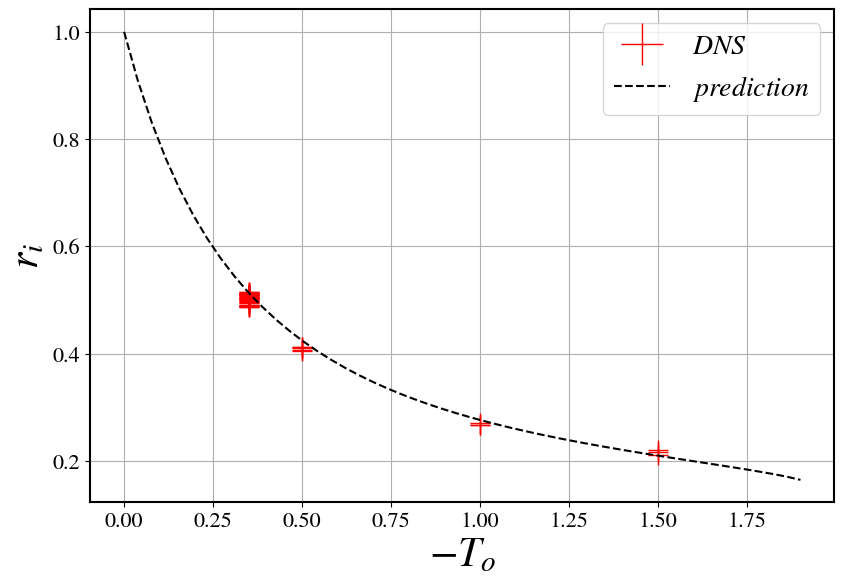}}% Images in 100% size
  \caption{Output interface radius $r_i$ as a function of the boundary condition of temperature $T_o$ obtained via (\ref{eq:apdx:ri}) and in the DNS.}
\label{fig:apdx1}
\end{figure}

\section{Table of results}\label{appB}

\tcw{Here we summarise the simulation outcomes for the parameter range considered.}
\FloatBarrier % prevents any earlier floats from sliding past this point

\begin{center}
%\captionof{table}{Summary of the simulations presented in this work, with their parameters $Ra$ and $S$, $Pr=0.01$ and $T_o/T_0=-0.35$, output measurements $\langle u_r'u_\phi'\rangle$, $\omega/N$, $kr_o$ and $\Lambda_1$ (see discussion about (\ref{eq:lbda1DNS})) and spatial resolution.}
\def~{\hphantom{0}}
\begin{tabular}{lllllll}
\hline
$Ra$        & $S$    & $\langle u_r'u_\phi'\rangle$ & $\overline{\omega}/N$ & $\overline{k}$    & $\Lambda_1$ & $(N_{r1}+ N_{r2}) \times N_\phi$ \\ [3pt]
$3.10^5$    & $1.98$ & $0.013445$                     & 0.486880   & 3.674618  & 0.029295    & $(32+32)\times 64$         \\
$10^{6}$    & $1.98$ & $0.013063$                     & 0.481802   & 3.841195  & 0.010201    & $(32+32)\times 64$         \\
$3.10^6$    & $1.98$ & $0.008850$                     & 0.461111   & 4.000100  & 0.006208    & $(64+64)\times 64$         \\
$10^7$      & $1.98$ & $0.004986$                     & 0.460197   & 4.111646  & 0.003513    & $(64+64)\times 128$        \\
$3.10^7$    & $1.98$ & $0.002712$                     & 0.451894   & 3.932658  & 0.002080    & $(64+64)\times 128$        \\
$10^{8}$    & $1.98$ & $0.001491$                     & 0.452695   & 3.975688  & 0.001154    & $(192+64)\times 256$       \\
$3.10^8$    & $1.98$ & $0.000695$                     & 0.465711   & 4.253614  & 0.000867    & $(192+64)\times 256$       \\
$10^{9}$    & $1.98$ & $0.000395$                     & 0.453485   & 4.639752  & 0.000591    & $(192+64)\times 256$       \\
$2.10^9$    & $1.98$ & $0.000277$                     & 0.459165   & 5.044808  & 0.000479    & $(192+64)\times 256$       \\
$3.10^9$    & $1.98$ & $0.000234$                     & 0.454688   & 5.297055  & 0.000429    & $(192+64)\times 256$       \\
$10^{10}$   & $1.98$ & $0.000138$                     & 0.429073   & 5.646715  & 0.000295    & $(192+64)\times 256$       \\
$3.10^{10}$ & $1.98$ & $0.000094$                     & 0.405666   & 6.259553  & 0.000211    & $(192+64)\times 256$       \\
$10^{11}$   & $1.98$ & $0.000053$                     & 0.392195   & 7.073847  & 0.000160    & $(192+96)\times 512$       \\
$10^{10}$   & $10$   & $0.000127$                     & 0.396151   & 7.690941  & 0.000754    & $(192+96)\times 256$       \\
$10^{10}$   & $19.8$ & $0.000108$                     & 0.397264   & 9.183230  & 0.001259    & $(192+96)\times 256$       \\
$3.10^{10}$ & $19.8$ & $0.000071$                     & 0.390387   & 10.51448 & 0.000875    & $(192+96)\times 256$       \\
$10^{11}$   & $19.8$ & $0.000051$                     & 0.413064   & 13.20753 & 0.000489    & $(256+128)\times 512$      \\
\hline
\end{tabular}
\label{tab:data}
\captionof{table}{Summary of the simulations presented in this work, with their parameters $Ra$ and $S$, $Pr=0.01$ and $T_o=-0.35$, output measurements $\langle u_r'u_\phi'\rangle$, $\overline{\omega}/N$, $\overline{k}$ and $\Lambda_1$ (see discussion about (\ref{eq:lbda1DNS})) and spatial resolution.}
\end{center}

%notes 2407205.txt
\end{appen}%\clearpage
%\clearpage

\bibliographystyle{jfm}
\bibliography{ref}

%Use of the above commands will create a bibliography using the .bib file. Shown below is a bibliography built from individual items.

%%%%%%%%%%%%% 1 by 1 works 
%\begin{thebibliography}{}
%\expandafter\ifx\csname natexlab\endcsname\relax
%\def\natexlab#1{#1}\fi
%\expandafter\ifx\csname selectlanguage\endcsname\relax
%\def\selectlanguage#1{\relax}\fi

%\bibitem[Aubert (2025)]{Aubert2025}
%{\sc Aubert, J.} 2025 {blabla}, {\it J. Fluid Mech.}, {\bf 5}, pp. 3-113-133.
%\end{thebibliography}

%% End of file `jfm.bib'.

\end{document}